%% file: VOADefQuant.tex
\documentclass[12pt,letterpaper]{article}
\usepackage{putex}
\usepackage{graphicx}
\usepackage{latexsym,amsmath,amsfonts,amssymb,amsthm}
\usepackage{bbm}
\usepackage{cite}
\usepackage{indentfirst}
\usepackage{booktabs,array}
\usepackage{placeins}
\usepackage{mathtools}
\usepackage{colonequals}
\usepackage{cancel}
\usepackage[mode=image|tex]{standalone}
\usepackage[dvipsnames]{xcolor}
\usepackage{tikz}
\usepackage{tikz-cd}
\usepackage{adjustbox}
\usepackage{genyoungtabtikz}
\usepackage{enumitem}
\usepackage{setspace}
\usetikzlibrary{decorations.markings}
\usetikzlibrary{decorations.text}
\usepackage{graphicx}
\usepackage[margin=10pt,font=small,labelfont=bf]{caption}
\usepackage{subcaption}
\usepackage{xcolor}
\usepackage{float}
\theoremstyle{definition}

\newtheorem{question}{Question}
\usepackage{microtype}
\usepackage{hyperref}
\hypersetup{unicode}
\hypersetup{linktoc = all}
\hypersetup{pdfborderstyle={/S/U/W 0.5}}
\hypersetup{linkbordercolor = gray}
\hypersetup{bookmarksnumbered}
\usepackage[T1]{fontenc}
\usepackage[utf8]{inputenc}
\usepackage{lmodern}

\setlist{itemsep=2pt plus 1pt minus 1pt, topsep=2pt plus 1pt minus 1pt}

\renewenvironment{thebibliography}[1]{%
\begin{oldthebibliography}{#1}%
\setlength\itemsep{-2mm}%
}%
{%
\end{oldthebibliography}%
}

\newcommand{\eg}{\textsl{e.g.\@}}

\newcommand{\ie}{\textsl{i.e.\@}}

\numberwithin{equation}{section}

\DeclareMathOperator{\Tr}{Tr}

\DeclareMathOperator{\rank}{rank}

\DeclareMathOperator{\re}{\mathbb{R}e}

\newcommand\qq{\mathbbmtt{Q}}

\begin{document}


\title{\begin{LARGE}
Deformation quantizations from vertex operator algebras\newline
\end{LARGE}}

\authors{Yiwen Pan$^1$ and Wolfger Peelaers$^2$
\medskip\medskip\medskip\medskip
 }

\institution{UU}{${}^1$
School of Physics, Sun Yat-Sen University, \cr
$\;\,$ Guangzhou, Guangdong, China}
\institution{Oxford}{${}^2$
Mathematical Institute, University of Oxford, Woodstock Road, \cr
$\;\,$ Oxford, OX2 6GG, United Kingdom}

\abstract{\begin{onehalfspace}{In this note we address the question whether one can recover from the vertex operator algebra associated with a four-dimensional $\mathcal N=2$ superconformal field theory the deformation quantization of the Higgs branch of vacua that appears as a protected subsector in the three-dimensional circle-reduced theory. We answer this question positively if the UV $R$-symmetries do not mix with accidental (topological) symmetries along the renormalization group flow from the four-dimensional theory on a circle to the three-dimensional theory. If they do mix, we still find a deformation quantization but at different values of its period.
}\end{onehalfspace}}

\preprint{}
\setcounter{page}{0}
\maketitle


{
\setcounter{tocdepth}{2}
\setlength\parskip{-0.7mm}
\tableofcontents
}

\input{./sections/introduction}
\input{./sections/ReviewVOADefQuant}
\input{./sections/fromVOAtoDefQuant}
\input{./sections/tests}

\section*{Acknowledgments}
The authors would like to thank Simone Giacomelli, Yongchao L\"u, Carlo Meneghelli, and especially Chris Beem for useful discussions and/or helpful suggestions. Y.P. is supported in part by the National Natural Science Foundation of China under Grant No. 11905301, the Fundamental Research Funds for the Central Universities under Grant No. 74130-31610023, the 100 Talents Program of Sun Yat-sen University under Grant No. 74130-18841207. The work of W.P. is partially supported by grant 494786 from the Simons Foundation.

\appendix
\input{./appendices/JacobiTheta}

\input{appendices/AKM}
\input{appendices/virasoro}
\input{./appendices/characters}

\clearpage

{
\bibliographystyle{utphys}
\bibliography{ref}
}

\end{document}

%% file: sections/introduction.tex

\section{Introduction and summary}
Conformal field theories famously possess a convergent, associative operator product algebra of local operators. Unfortunately, an analytic handle on this algebraic structure appears well out of reach in dimensions larger than two. If the symmetry algebra of the theory additionally includes supersymmetry, however, one can attempt to perform a cohomological truncation with respect to one of its nilpotent supercharges and aim to study the necessarily simpler, still associative algebra of cohomology classes instead. This approach has met with great success. In the seminal paper \cite{Beem:2013sza}, it was shown that a construction of precisely this type defines a correspondence between four-dimensional $\mathcal N=2$ superconformal field theories (SCFTs) and vertex operator algebras (VOAs):\footnote{This SCFT/VOA correspondence has resulted in a large number of interesting results. For example, many SCFT/VOA pairs have been identified in, \eg{}, \cite{Cordova:2015nma,Xie:2016evu,Song:2017oew,Buican:2017fiq,Choi:2017nur,Creutzig:2018lbc,Xie:2019yds}, and, more remarkably, the correspondence has inspired the construction of novel vertex operator algebras \cite{Beem:2014rza,Lemos:2014lua,Nishinaka:2016hbw,Bonetti:2018fqz,Arakawa:2018egx}. It has also served as a useful tool to acquire new insights in SCFTs and VOAs, see for example \cite{Buican:2015ina,Liendo:2015ofa,Cecotti:2015lab,Lemos:2015orc,Song:2016yfd,Buican:2016arp,Fredrickson:2017yka,Beem:2017ooy,Beem:2018duj,Pan:2019bor,Beem:2019tfp,Oh:2019bgz,Jeong:2019pzg,Dedushenko:2019yiw,Xie:2019zlb,Beem:2019snk,Oh:2019mcg,Watanabe:2019ssf,Xie:2019vzr}.}\textsuperscript{,}\footnote{Also six-dimensional $\mathcal N=(2,0)$ SCFTs participate in a truncation to vertex operator algebras \cite{Beem:2014kka}.}
\begin{equation}\label{VOAcorr}
\mathbb V: \{\text{4d $\mathcal N=2$ SCFTs} \} \longrightarrow \{\text{VOAs}\}\;.
\end{equation}
Similarly, three-dimensional $\mathcal N=4$ superconformal field theories can be mapped to topological algebras \cite{Chester:2014mea,Beem:2016cbd}. What's more, these topological algebras have been shown to be a deformation quantization of the ring of holomorphic functions over the Higgs branch of vacua $\mathcal M_H$ \cite{Beem:2016cbd}.\footnote{A similar construction exists for the Coulomb branch of vacua.} Three-dimensional $\mathcal N=4$ superconformal symmetry dictates that the Higgs branch is a hyperk\"ahler cone, so we have
\begin{equation}\label{DQcorr}
\mathbb{DQ}: \{\text{3d $\mathcal N=4$ SCFTs} \} \longrightarrow \{\text{deformation quantizations of hyperk\"ahler cones}\}\;.
\end{equation}
It is important to remark that the image of $\mathbb{DQ}$ is a $\mathbb C^*$-equivariant, even deformation quantization satisfying two additional properties. The first one states that the deformed multiplication truncates sooner than $\mathbb C^*$-equivariance demands, while the second one implements the physical condition of unitarity. The aim of this note is to investigate the following question
\begin{question}\label{question}
Can one recover from the vertex operator algebra associated with a four-dimensional $\mathcal N=2$ superconformal field theory $\mathcal T_{4d}$ the deformation quantization corresponding to the dimensional reduction $\mathcal T_{3d}$ of that SCFT?
\end{question}

\begin{figure}
\centering
\begin{tikzpicture}                    
\node(T)  at (0,0) {$\mathcal T_{4d}$};
\node(VOA)at (3,0) {$\mathbb V[\mathcal T_{4d}]$};
\node(HB) at (0,-2.5) {$\mathcal M_H[\mathcal T_{4d}]$};
\draw[|->] (T) -- (VOA) node[midway, above]{$\mathbb V$};
\draw[|->] (T) -- (HB) node[midway,left] {HB};
\draw[|->] (VOA) -- (HB) node[midway,above,sloped] {Assoc. Var.};
\draw[|->,dotted,postaction={decorate,decoration={raise=-2.5ex,text along path,text align=center,text={free field constr.}}}] (HB) to [out=0,in=-90] (VOA);

\node(T3d)  at (0,-6) {$\mathcal T_{3d}$};
\node(HB3d) at (0,-3.5) {$\mathcal M_H[\mathcal T_{3d}]$};

\node(DQ3d) at (3,-6) {$\mathbb {DQ}[\mathcal T_{3d}]$};

\node(equal)  at (0,-3) {\rotatebox{90}{$\,\equiv$}};

\draw[|->] (T3d) -- (HB3d) node[midway,left] {HB};
\draw[|->] (T) to [out=210,in=150] node[midway,left]{dim. red.} (T3d);
\draw[|->] (T3d) -- (DQ3d) node[midway,below] {$\mathbb {DQ}$};

\draw[|->] (VOA) to [out=-30,in=30] node[midway,right]{?} (DQ3d);
\draw[|->] (DQ3d) -- (HB3d) node[midway,right] {classical limit};

\end{tikzpicture}
\caption{\label{figinterconn} Interconnections between various objects associated with the four-dimensional theory $\mathcal T_{4d}$ and its three-dimensional dimensional reduction $\mathcal T_{3d}$.}
\end{figure}
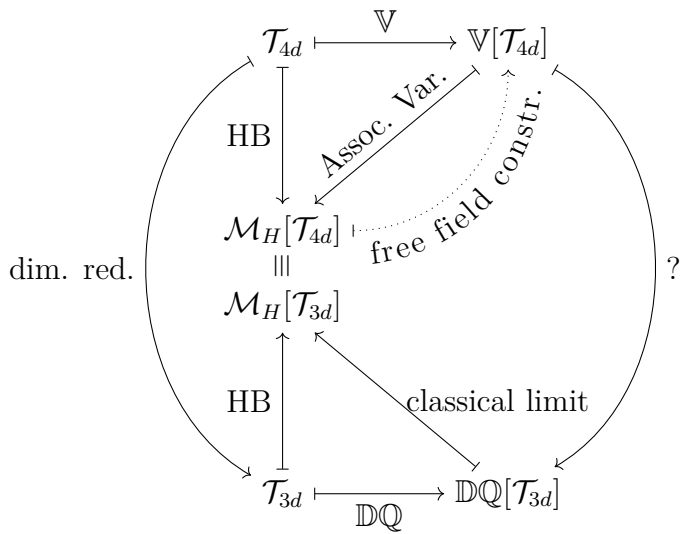

An obvious first clue that this may be possible is the common lore that the Higgs branch of vacua remains unchanged under dimensional reduction: $\mathcal M_H[\mathcal T_{4d}] \equiv \mathcal M_H[\mathcal T_{3d}]$. Furthermore, in the past few years it has become clear that Higgs branch physics is intricately encoded in $\mathbb V[\mathcal T_{4d}]$. Indeed, given $\mathbb V[\mathcal T_{4d}]$, one can (conjecturally) recover $\mathcal M_H[\mathcal T_{4d}]$ by taking the so-called associated variety \cite{Beem:2017ooy}, while free-field realizations mirroring the effective field theory description of the SCFT in a Higgs vacuum allow one to reconstruct the vertex operator algebra itself \cite{Beem:2019tfp,Beem:2019snk}. All in all, we thus have interconnections as in figure \ref{figinterconn}, and we aim to fill in the question mark in this figure.

Our strategy to answer question \ref{question} was suggested in our previous paper \cite{Pan:2019bor} and is as follows. An important entry of the SCFT/VOA dictionary states that the vacuum character of the VOA equals a particular limit of the superconformal index of the SCFT---the so-called Schur limit. This index can be defined as the partition function of $\mathcal T_{4d}$ placed supersymmetrically on a manifold with topology $S^3\times S^1$. In \cite{Pan:2019bor,Dedushenko:2019yiw}, this equality of the torus partition function of the vertex operator algebra and the Schur index of the SCFT was made explicit using supersymmetric localization techniques. Indeed, for Lagrangian theories it was proved that the path integral can be localized to a slice of field configurations describing the VOA living on a torus $\mathbb T^2 \subset S^3\times S^1$. What's more, the computation was extended to efficiently compute torus correlation functions of fields of the vertex operator algebra. Similarly, in \cite{Dedushenko:2016jxl} a supersymmetric localization computation of correlation functions defining the topological algebra was performed on the three-sphere $S^3$.\footnote{See \cite{Dedushenko:2017avn,Dedushenko:2018icp} for a localization computation of the deformation quantization of the Coulomb branch.} The key observation is now that, roughly speaking, upon dimensionally reducing the former computation along the circle $S^1$, one lands on the latter. Moreover, although localization techniques are not available for non-Lagrangian theories, there is no \textit{a priori} obstruction to formalizing this procedure and to apply it to non-Lagrangian theories as well. It amounts to, still roughly speaking, taking the high-temperature limit $\beta\rightarrow 0$ of torus correlation functions of the vertex operator algebra. The qualifiers ``roughly speaking'' refer to various subtleties. First of all, the fields of the vertex operator are strictly larger in number than those participating in the deformation quantization. To remove the spurious fields, we accompany the high-temperature limit with a particular rescaling of the fields of the VOA. In detail, if $a(z)$ is a field of conformal weight $h_a$, we rescale as $a(z) \rightarrow \beta^{h_a} a(z)$. Second, the results on the three-sphere suffer from operator mixing, which can be disentangled by a Gram-Schmidt process that aims to diagonalize the matrix of two-point functions. Third, by moving outside the realm of Lagrangian theories, we have opened the door for the IR $R$-symmetries of the three-dimensional SCFT being realized as a mixture of the four-dimensional UV $R$-symmetries and accidental (topological) symmetries emerging along the renormalizaton group flow from the four-dimensional UV SCFT to the three-dimensional IR SCFT. This type of mixing leaves an imprint on the three-dimensional theory on the three-sphere that takes the form of imaginary Fayet-Iliopoulos parameters \cite{Buican:2015hsa}.\footnote{Recall that Fayet-Iliopoulos parameters are mass parameters for topological symmetries and that on the three-sphere imaginary parts of mass parameters encode $R$-symmetry admixtures \cite{Jafferis:2010un}.} These directly affect the deformation quantization and ultimately spell an end to any hopes of answering our question in the positive in all generality---at least not in any obvious manner. Only if such mixing does not occur will our procedure produce $\mathbb {DQ}[\mathcal T_{3d}]$ from $\mathbb V[\mathcal T_{4d}]$. In all other cases, we do find a deformation quantization, but at different values of its parameters than $\mathbb {DQ}[\mathcal T_{3d}]$.

This note is organized as follows. In section \ref{review}, we briefly review the SCFT/VOA and the SCFT/deformation quantization correspondences. In section \ref{proposal} we present in detail our proposal for recovering $\mathbb {DQ}[\mathcal T_{3d}]$ from $\mathbb {V}[\mathcal T_{4d}]$. Section \ref{tests} contains the proof of our proposal in Lagrangian theories and section \ref{nonlagrtests} various tests and examples for non-Lagrangian theories. We collect various useful functions and their properties in appendix \ref{specialfuctions}. Appendix \ref{AKM} contains explicit results for torus correlation functions of current algebras and their resulting deformation quantizations. In appendix \ref{app:stresstensor}, we present an analysis of Virasoro torus correlation functions based on Ward identities. Finally, appendix \ref{characters} collects characters of various vertex operator algebras studied in sections \ref{tests} and \ref{nonlagrtests}.

\medskip

\textit{Note added:} when this note was being finalized, the paper \cite{Dedushenko:2019mzv} appeared on the arXiv, which addresses the same question.

%% file: sections/ReviewVOADefQuant.tex

\section{Algebraic structures in 4d and 3d SCFTs}\label{review}
The aim of this note is to relate vertex operator algebraic structures one can carve out in four-dimensional $\mathcal N=2$ superconformal field theories to topological algebras appearing in their three-dimensional $\mathcal N=4$ superconformal dimensional reduction. In this section we start by briefly reviewing how these algebras emerge.  

\subsection{SCFT/VOA correspondence}
Four-dimensional $\mathcal N=2$ superconformal field theories, \ie{}, quantum field theories whose symmetry algebra contains the superalgebra $\mathfrak{su}(2,2|2)$, take part in a remarkable correspondence with vertex operator algebras \cite{Beem:2013sza,Beem:2014rza}: 
\begin{equation}
\mathbb V: \{\text{4d $\mathcal N=2$ SCFTs} \} \longrightarrow \{\text{VOAs}\}\;.
\end{equation}
The vertex operator algebra arises as a cohomological reduction of the operator product algebra of local operators of the SCFT. The cohomology is computed with respect to a nilpotent supercharge $\qq\in \mathfrak{su}(2,2|2)$ that takes the schematic form of the sum of a Poincar\'e supercharge and a special conformal supercharge: $\qq=Q+S$. At the origin of space, harmonic representatives of $\qq$-cohomology classes are characterized by a linear relation among their quantum numbers. They were dubbed ``Schur operators'', because they are precisely the operators counted by the Schur limit of the superconformal index \cite{Gadde:2011uv}. Concretely,
\begin{equation}\label{SchurOpsDefinition}
\mathcal O(0)\ \text{is a Schur operator} \qquad \Longleftrightarrow \qquad E - 2R - j_1 - j_2 = 0\;,
\end{equation}
where $E$ denotes the conformal dimension of the operator $\mathcal O$, $(j_1,j_2)$ its $\mathfrak{su}(2)_1\times \mathfrak{su}(2)_2$ rotational quantum numbers, and $R$ the value of its $SU(2)_R$ Cartan charge. Unitarity of the theory implies that these operators automatically satisfy $r+j_1-j_2=0$, where $r$ denotes the $U(1)_r$ charge. They can be transported away from the origin while remaining in cohomology by $\qq$-closed translation operators. It is an algebraic fact that only motion in a complex plane $\mathbb C_{[z,\bar z]}$ is possible, and, moreover, that the (twisted) translation in the $\bar z$-direction is $\qq$-exact.\footnote{The subalgebra of the four-dimensional conformal algebra $\mathfrak{so}(4,2)$ that preserves the complex plane $\mathbb C_{[z,\bar z]}$ set-wise is $\mathfrak{sl}(2)_z\times\overline{\mathfrak{sl}(2)}_{\bar z}\times \mathfrak{u}(1)_{\perp}$. The $\mathfrak{sl}(2)$ factors generate the standard M\"obius transformations on the respective complex coordinate and $\mathfrak{u}(1)_{\perp}$ contains rotations orthogonal to $\mathbb C_{[z,\bar z]}$. While the $\qq$-closed translation of $z$ is simply generated by $L_{-1}$, the $\qq$-exact translation of the coordinate $\bar z$ is generated by $\bar L_{-1}+R^-$, where $R^-$ is the $\mathfrak{su}(2)_R$ lowering operator. Due to the presence of the $SU(2)_R$ generator, one often speaks of ``twisted translation.''} Therefore, cohomology classes depend only on the holomorphic coordinate $z$. The operator algebra of the four-dimensional SCFT endows these holomorphic cohomology classes with the structure of a vertex operator algebra. Note that the (holomorphic) conformal weight of a Schur operator is measured by the eigenvalue of $L_0$, the Cartan generator of the $\mathfrak{sl}(2)_z$ algebra of M\"obius transformations of $z$. Upon embedding this algebra in the four-dimensional conformal algebra, one finds on any Schur operator $\mathcal O(0)$ that
\begin{equation}
[L_0,\mathcal O(0)] = h\, \mathcal O(0)\;, \qquad \text{with} \qquad h = \frac{E + j_1 + j_2}{2} = E-R\;.
\end{equation}
To arrive at the second equality, we used the relation among quantum numbers of \eqref{SchurOpsDefinition}.

The set of Schur operators is non-empty for any local SCFT. Indeed, the four-dimensional stress-energy tensor, which is guaranteed to exist in a local quantum field theory, resides in a superconformal multiplet of $\mathfrak{su}(2,2|2)$ that contains a Schur operator -- a certain component of the Noether current of the $SU(2)_R$ symmetry. Its twisted translation defines a holomorphic cohomology class whose operator product expansions identify it as the Virasoro stress tensor of the VOA. The Virasoro central charge $c_{2d}$ is related to the Weyl anomaly coefficient $c_{4d}$ of the above-lying superconformal field theory by the universal relation
\begin{equation}
c_{2d} = -12 c_{4d}\;.
\end{equation}

A collection of additional Schur operators are related to the Higgs branch geometry of the SCFT.\footnote{By no means do the operators associated with Higgs branch chiral ring operators exhaust the full set of Schur operators. See \cite{Beem:2013sza} for a complete dissection of the body of Schur operators.} Four-dimensional $\mathcal N=2$ superconformal field theories come equipped with a (possibly empty) Higgs branch of vacua $\mathcal M_H$, which can be defined as the branch of the moduli space of vacua preserving the $U(1)_r$ symmetry. Its geometric structure is that of a hyperk\"ahler cone.\footnote{Recall that a hyperk\"ahler cone features three complex structures $J_i$, $i=1,2,3$, that anticommute with one another, satisfy $J_i^2=-1$ and $J_1J_2=J_3$, and are compatible with the metric $g(X,Y)=g(J_iX,J_iY)$. The three complex structures are rotated as a triplet by an $SU(2)$ isometry. Their associated K\"ahler forms $\omega_i$ are closed $d\omega_i=0$. The conical structure further implies that the metric takes the form $ds^2 = dr^2 + r^2 ds^2_{\text{base}}$.} Its coordinate ring $\mathbb C[\mathcal M_H]$, \ie{}, the ring of holomorphic functions over the cone, can be identified with the Higgs branch chiral ring.\footnote{\label{footnoteCoordRing}To define the coordinate ring $\mathbb C[\mathcal M_H]$ of the hyperk\"ahler cone $\mathcal M_H$, we have singled out one of the complex structures, say $J_3$. Thanks to the $SU(2)$ isometry acting on the complex structures, all choices are equivalent. The other two K\"ahler forms organize themselves into a holomorphic symplectic form $\Omega^{(2,0)} = \omega_1+i\omega_2$, turning the coordinate ring $\mathbb C[\mathcal M_H]$ into a commutative, associative Poisson algebra. Note that the $SU(2)$ isometry of the hyperk\"ahler cone is identified with the R-symmetry $SU(2)_R$. The choice of complex structure is correlated with the Cartan decomposition of this $SU(2)$ group. Finally, for future purposes, it is worth mentioning that the coordinate ring admits a $\mathbb C^*$ grading, which is the complexification of the dilatation symmetry with the $U(1)_R\subset SU(2)_R$ Cartan subalgebra.} The elements of the latter are defined as
\begin{equation}\label{HBOpsDefinition4d}
\mathcal O(0)\ \text{is a 4d $\mathcal N=2$ Higgs branch chiral ring operator} \qquad \Longleftrightarrow \qquad E = 2R\;,
\end{equation}
and their multiplication is obtained from the coincident limit of their standard operator product expansion. Note that Higgs branch chiral ring operators are automatically Lorentz scalars $j_1=j_2=0$ and $U(1)_r$ neutral, and thus are special instances of Schur operators. What's more, one can prove that all Higgs branch chiral ring operators give rise to Virasoro primary operators in the vertex operator algebra and that the generators of the ring are strong generators of the VOA.

Of particular importance in this class of Schur operators are the moment map operators associated with flavor symmetries $G_F$ of the SCFT, or, equivalently, with hyperk\"ahler isometries of the Higgs branch. Their twisted-translated counterparts satisfy meromorphic operator product expansions defining an affine current algebra $\hat{\mathfrak g}_F$. The level $k_{2d}$ of this Kac-Moody algebra is related to the four-dimensional flavor central charge $k_{4d}$ via
\begin{equation}
k_{2d} = -\frac{1}{2}k_{4d}\;.
\end{equation}

It is clear from the preceding discussion that the vertex operator algebra is intimately related to the geometry of the Higgs branch of vacua. This relationship was fleshed out and made precise in \cite{Beem:2017ooy} (see also \cite{arakawa2012remark,Arakawa:2015jya}). In that paper, it was conjectured that the Higgs branch can be recovered (as a holomorphic symplectic variety) from the vertex operator algebra as the so-called associated variety of the VOA. This variety is defined as the spectrum of the ring one obtains by performing a certain quotient of the ring of VOA operators whose multiplication is the standard normal ordered product. \textit{Vice versa}, strong evidence has been obtained in \cite{Beem:2019tfp,Beem:2019snk} that the VOA can be given a (generalized) free-field construction mirroring the effective field theory description of the SCFT in a Higgs vacuum.\footnote{See also \cite{Bonetti:2018fqz} for a different kind of free-field realizations.} Schematically, the situation is thus as follows
\begin{center}

\begin{tikzpicture}
                    
    \node(T)  at (0,0) {$\mathcal T_{4d}$};
    \node(VOA)at (3,0) {$\mathbb V[\mathcal T_{4d}]$};
    \node(HB) at (0,-2.5) {$\mathcal M_H[\mathcal T_{4d}]$};
    \draw[|->] (T) -- (VOA) node[midway, above]{$\mathbb V$};
    \draw[|->] (T) -- (HB) node[midway,left] {HB};
    \draw[|->] (VOA) -- (HB) node[midway,above,sloped] {Assoc. Var.};
    \draw[|->,dotted,postaction={decorate,decoration={raise=-2.5ex,text along path,text align=center,text={free field constr.}}}] (HB) to [out=0,in=-90] (VOA);
    \end{tikzpicture}

\end{center}

We conclude this subsection by expressing in formulae the statement made implicitly above that the vacuum character of the vertex operator algebra $\chi_0(q)$ equals the Schur limit of the superconformal index $I_S(q)$
\begin{equation}\label{vacchar=index}
\chi_0(q) \colonequals \text{STr}\ q^{L_0-c_{2d}/24} \equiv q^{c_{4d}/2}\ \Tr_{\mathcal H(S^3)} (-1)^F q^{E-R} \equalscolon I_{S}(q)\;.
\end{equation}
Here $\text{STr}$ denotes the supertrace over the space of states of the VOA and $\mathcal H(S^3)$ is the Hilbert space of states on the three-sphere of the four-dimensional SCFT. In the trace over the latter, many cancellations take place and ultimately only Schur operators contribute. As we will review in more detail below, both sides of this identification can be decorated with (Schur) operator insertions. For Lagrangian theories, the Schur index dressed with additional insertions can be computed explicitly using supersymmetric localization techniques as a correlation function of the four-dimensional SCFT placed on $S^3\times_q S^1$ \cite{Pan:2019bor} (see also \cite{Dedushenko:2019yiw}).

\subsection{SCFT/deformation quantization correspondence}\label{subsecdefquant}
Both the Coulomb branch $\widetilde{\mathcal M}_C$ and Higgs branch $\widetilde{\mathcal M}_H$ of vacua of three-dimensional $\mathcal N=4$ superconformal field theories are hyperk\"ahler cones. They are singled out as the loci of the moduli space of vacua kept invariant by the R-symmetry algebras $SU(2)_H$ and $SU(2)_C$, respectively.\footnote{Recall that three-dimensional $\mathcal N=4$ superconformal field theories are defined by the requirement that their symmetry algebra contain the superalgebra $\mathfrak{osp}(4|4,\mathbb R)$. Its bosonic subalgebra is given by $(\mathfrak{su}(2)_C\oplus\mathfrak{su}(2)_H)\times \mathfrak{sp}(4,\mathbb R)$, where one recognizes the R-symmetry algebras and the three-dimensional conformal algebra $ \mathfrak{sp}(4,\mathbb R) \simeq \mathfrak{so}(3,2)$.} In this note we will focus on the Higgs branch. Higgs branch chiral ring operators of three-dimensional $\mathcal N=4$ SCFTs are characterized as follows:
\begin{equation}\label{HBOpsDefinition3d}
\widetilde{\mathcal O}(0)\ \text{is a 3d $\mathcal N=4$ Higgs branch chiral ring operator} \qquad \Longleftrightarrow \qquad \widetilde E = R_H\;,
\end{equation}
where $\widetilde E$ denotes the conformal dimension and $R_H$ the $SU(2)_H$ Cartan charge of the operator $\widetilde{\mathcal O}(0)$. They are Lorentz scalars and singlets under $SU(2)_C$. As before, the Higgs branch chiral ring is identified with the coordinate ring $\mathbb C[\widetilde{\mathcal M}_H]$. What's more, trivially dimensionally reducing a four-dimensional $\mathcal N=2$ SCFT with Higgs branch $\mathcal M_H$ and flowing towards the infrared results in a three-dimensional $\mathcal N=4$ SCFT with Higgs branch $\widetilde{\mathcal M}_H = \mathcal M_H$.

Very similarly to the SCFT/VOA correspondence of the previous subsection, three-dimensional $\mathcal N=4$ superconformal field theories admit a cohomological truncation to a one-dimensional topological algebra, \ie{}, an associative algebra additionally endowed with an evaluation map \cite{Chester:2014mea,Beem:2016cbd}. As was uncovered in \cite{Beem:2016cbd}, this algebra is a noncommutative deformation of the coordinate ring $\mathbb C[\widetilde {\mathcal M}_H]$.\footnote{Indeed, harmonic representatives of cohomology classes at the origin are characterized by the requirement that their quantum numbers satisfy $\widetilde E = R_H$, \ie{}, that they are Higgs branch chiral ring operators. While remaining in cohomology, these operators can only be (twisted) translated away from the origin along a line, and the coordinate dependence along this line is exact. Thus, only the ordering of the twisted-translated cohomology classes along the line is retained. Their algebraic properties descend from the three-dimensional operator product algebra: they form an associative, but not necessarily commutative, algebra, and taking vacuum expectation values provides the above-mentioned evaluation map.} Moreover, it was shown in that same paper that the leading term of this deformation is determined by the Poisson bracket: the topological algebra thus defines a deformation quantization:\footnote{Recall from footnote \ref{footnoteCoordRing} that the coordinate ring of a hyperk\"ahler cone is naturally a Poisson algebra.}
\begin{equation}
\mathbb{DQ}: \{\text{3d $\mathcal N=4$ SCFTs} \} \longrightarrow \{\text{deformation quantizations of hyperk\"ahler cones}\}\;.
\end{equation}

Various other properties of this deformation have been established in \cite{Beem:2016cbd} and can be summarized as follows. Let $f \in \mathcal A_p, g\in \mathcal A_q$ be elements of the $p$th and $q$th $\mathbb C^*$-graded component of the coordinate ring $\mathcal A = \mathbb C[\widetilde {\mathcal M}_H]$. In other words, $f$ and $g$ are holomorphic functions over the Higgs branch that correspond to Higgs branch chiral ring operators of $SU(2)_H$ charge $R_H = \frac{p}{2}$ and $\frac{q}{2}$ respectively. Then the multiplication of their twisted-translated cohomology classes ordered along the line, which we denote by $\star$, reads
\begin{equation}\label{generalformstarproduct}
f\star g = f\cdot g  + \frac{\zeta}{2} \{f,g \}_{PB} + \sum_{k=2}^{\lfloor \frac{p+q}{2}\rfloor}\zeta^k C^k(f,g)\;,
\end{equation}
where $f\cdot g$ is simply the multiplication in the ring $\mathcal A$ and $\{f,g \}_{PB}$ denotes the Poisson bracket. Here $\zeta$ is an immaterial book-keeping device that keeps track of one unit of $SU(2)_H$-charge. Moreover, the $\star$-product satisfies the following properties:
\begin{enumerate}
\item \makebox[4.9cm][l]{associativity of the OPE}  $\quad\Longrightarrow\quad$ \makebox[5.6cm][l]{$f\star(g\star h) = (f\star g)\star h$}(associativity)
\item  \makebox[4.9cm][l]{$SU(2)_H$-charge conservation} $\quad\Longrightarrow\quad$ \makebox[5.6cm][l]{$C^k:\mathcal A_p \otimes \mathcal A_q \rightarrow \mathcal A_{p+q-2k}$}($\mathbb C^*$-equivariance)
\item  \makebox[4.9cm][l]{$SU(2)_H$ selection rules} $\quad\Longrightarrow\quad$ \makebox[5.6cm][l]{$C^k(f,g)=0 \text{ for } k>\text{min}(p,q)$}(truncation)
\item  \makebox[4.9cm][l]{symmetry properties OPE} $\quad\Longrightarrow\quad$ \makebox[5.6cm][l]{$C^k(f,g)=(-1)^k C^k(g,f)$}(evenness)
\end{enumerate}
A fifth property is related to unitarity. Let $\rho$ be a rotation over $\pi$ in $SU(2)_H$ followed by complex conjugation, then for $f_1,f_2\in \mathcal A_p$
\begin{itemize}
\item[5.]  \makebox[2cm][l]{unitarity} $\quad\Longrightarrow\quad$ \makebox[5.6cm][l]{$\theta(f_1,f_2) \colonequals C^p(\rho(f_1),f_2)$ is a positive definite Hermitian form}
\end{itemize}
The general form \eqref{generalformstarproduct} of the $\star$-product together with properties 1, 2 (and 4) are the mathematical definition of an (even) $\mathbb C^*$-equivariant deformation quantization of the Poisson algebra $(\mathcal A,\cdot,\{.,.\}_{PB})$. As explained in \cite{Beem:2016cbd}, properties 3 and 5 are gauge fixing conditions for an infinite-dimensional group of gauge equivalences. They were conjectured to be perfect in \cite{Beem:2016cbd}. All in all, the deformation quantization thus depends on a finite number of intrinsic parameters, the so-called period of the quantization. This period is in one-to-one correspondence with elements of $H^2(\widetilde{ { m}}_H, \mathbb C)/W$, where $\widetilde{{ m}}_H$ is a smooth symplectic resolution of $\widetilde{ {\mathcal M}}_H$ and $W$ the Namikawa-Weyl group \cite{Braden:2014cb,Braden:2014iea}. In more physical terms, the Namikawa-Weyl group is simply the Weyl group of the Coulomb branch flavor symmetry of the theory, while $H^2(\widetilde{ {m}}_H, \mathbb C)$ can be identified with the space of real mass parameters of topological Coulomb branch symmetries, \ie{}, Fayet-Iliopoulos terms, one could in principle turn on in the theory \cite{Beem:2016cbd}.\footnote{We do not, however, actually turn on these parameters as they would break conformal invariance. The identification of spaces is only meant to clarify its definition.} Exciting progress towards proving this conjecture has recently been made in the mathematics literature \cite{Etingof:2019guc}. 

%% file: sections/fromVOAtoDefQuant.tex

\section{From VOA to deformation quantization}\label{proposal}
As we have explained in the previous section, both the vertex operator algebras associated with four-dimensional $\mathcal N=2$ SCFTs and the deformation quantizations corresponding to three-dimensional $\mathcal N=4$ SCFTs are intimately related to the geometry of their Higgs branches of vacua -- the latter in an obvious manner, as they are the quantizations of the coordinate ring of the Higgs branch, while the former in a slightly less manifest manner. Moreover, it is a well-known fact that the Higgs branch of vacua remains invariant under (trivial) dimensional reduction. A natural question to ask is then if one can fill in the question mark in the following diagram:
\begin{center}
\begin{tikzpicture}
    \node(T)  at (0,0) {$\mathcal T_{4d}$};
    \node(VOA)at (2.5,0) {$\mathbb V[\mathcal T_{4d}]$};
    \node(T3d) at (0,-2) {$\mathcal T_{3d}$};
    \node(DQ) at (2.5,-2) {$\mathbb {DQ}[\mathcal T_{3d}]$};
    \draw[|->] (T) -- (VOA) node[midway, above]{$\mathbb V$};
    \draw[|->] (T) -- (T3d) node[midway,left] {dim. red.};
    \draw[|->] (T3d) -- (DQ) node[midway, above]{$\mathbb {DQ}$};
    \draw[|->] (VOA) -- (DQ) node[midway,right] {?};
    \end{tikzpicture}
\end{center}
\noindent In this section, we will attempt to answer this question by formalizing the dimensional reduction of the four-dimensional theory $\mathcal T_{4d}$ placed supersymmetrically on the (warped) product-manifold $S^3\times_q S^1$. The partition function on this background computes the Schur limit of the superconformal index: 
\begin{equation}\label{index=partfnct}
Z_{S^3\times_q S^1}[\mathcal T_{4d}] = I_S^{(\mathcal T_{4d})}(q). 
\end{equation}
In other words, using \eqref{vacchar=index}, it computes the torus partition function of the vertex operator algebra associated with the SCFT. The relevant torus $\mathbb T^2 \subset S^3\times_q S^1$ arises as the point-wise fixed set of a certain spatial $U(1)$ rotation; its complex structure $\tau$ is determined by $q = e^{2\pi i \tau}$. Moreover, it was shown in \cite{Pan:2019bor} (see also \cite{Dedushenko:2019yiw}) that one can enrich the computation with (twisted-translated) Schur operators inserted at points on said torus $\mathbb T^2$.\footnote{A similar result with vertex operator algebra insertions restricted to a two-sphere can be obtained on the four-sphere, see \cite{Pan:2017zie}.} The resulting $S^3\times_q S^1$ correlators equal torus correlation functions of the associated fields of $\mathbb V[\mathcal T_{4d}]$. What's more, for Lagrangian SCFTs one can leverage supersymmetric localization techniques to arrive at an effective and efficient computational method of these correlation functions.

If $\mathcal T_{3d}$ admits a UV Lagrangian description in terms of vector and hypermultiplets, it was shown in \cite{Dedushenko:2016jxl} that the correlation functions defining the one-dimensional topological algebra associated with (the Higgs sector of) $\mathcal T_{3d}$ are accessible via a localization computation on the three-sphere. Note however that operators on the three-sphere may mix, see \cite{Gerchkovitz:2016gxx,Dedushenko:2016jxl}. The correct flat-space basis can be determined by diagonalizing the matrix of two-point functions. As we suggested in \cite{Pan:2019bor}, and will show in detail in section \ref{tests}, dimensionally reducing the Lagrangian setup on $S^3\times_q S^1$ along $S^1$ allows one to make direct contact with these computations. Moreover, by observing that this $S^1$-reduction is nothing but the high-temperature limit $\tau\rightarrow +i 0$, \ie{}, $\beta \colonequals -i \tau \rightarrow +0$, of the torus correlation functions of $\mathbb V[\mathcal T_{4d}]$, it can be phrased purely vertex operator algebraically. This formulation is helpful because the VOA is often known even in the absence of a Lagrangian description of $\mathcal T_{4d}$. 

Simply considering the high-temperature limit of torus correlators cannot be the complete story, since, as explained above, the set of Schur operators is strictly bigger than the collection of Higgs branch chiral ring operators. Taking a cue from the notion of contracting algebras, we propose a rescaling of the four-dimensional Schur operators by powers of $\beta$ that ensures that only correlation functions of Higgs branch chiral ring operators survive in the $\beta\rightarrow 0$ limit. Finally, one should observe that the $R$-symmetries of the three-dimensional theory in the IR may be realized as a mixture of the four-dimensional UV $R$-symmetries and accidental (topological) symmetries. A simple criterion for when this must be the case was put forward in \cite{Buican:2015hsa}, namely if the Coulomb branch chiral ring has generators of $U(1)_r$ charges that are not quantized in half-integer units. Such mixing is encoded in the three-sphere partition function as imaginary parts of the Fayet-Iliopoulos parameters. Such parameters directly modify the period of the resulting quantization, thus taking us away from the specific superconformal deformation quantization $\mathbb{DQ}(\mathcal{T}_{3d})$.

\subsection{Torus correlation functions}
We start by putting the vertex operator algebra $\mathbb V[\mathcal T_{4d}]$ on the torus $\mathbb T^2 = \mathbb C/(\mathbb Z + \tau \mathbb Z)$. The modular parameter $\tau$ takes values in the upper half plane $\mathbb H$; it is acted on by the modular group $PSL(2,\mathbb Z)$ in the usual fashion: $\tau\mapsto \frac{a \tau +b}{c\tau +d}$, where $\left(\begin{smallmatrix} a & b \\ c & d\end{smallmatrix}\right)\in PSL(2,\mathbb Z)$. The quantities of our interest are normalized torus correlation functions. Concretely, for primary fields $\mathcal O_I$ of conformal weight $h_I$, one finds\footnote{\label{footnoteperiodicity}In writing this equation we have slightly abused notation by giving the same name to both the operator on $\mathbb T^2$ and $\mathbb C$. Their coordinate-dependence distinguishes them. Also note that $\langle \ldots \rangle_{\mathbb T^2}$ denotes the correlation function of fields that are periodic along the temporal cycle, as implemented by the supertrace, and (anti)periodic along the spatial cycle depending on their (two-dimensional) spin.}
\begin{equation}
\langle \mathcal O_1(z_1) \mathcal O_2(z_2)\ldots  \mathcal O_n(z_n)  \rangle_{\mathbb T^2} = \frac{\prod_{I=1}^n (2\pi iw_I)^{h_{I}}}{\chi_0(q)} \text{Str}\ \mathcal O_1(w_1) \mathcal O_2(w_2)\ldots  \mathcal O_n(w_n) \ q^{L_0-\frac{c_{2d}}{24}}\;,
\end{equation}
where $w_j = e^{2\pi i z_j}$ relates the coordinates $z_j$ on the torus to coordinates $w_j$ on the plane. The normalization is provided by the unnormalized one-point function of the identity operator $\mathcal O_{\mathbbm 1} = \mathbbm 1$, in other words, the vacuum character:
\begin{equation}
\langle\mathbbm 1\rangle_{\mathbb T^2, \text{unnorm.}} = \text{Str}\ q^{L_0-\frac{c_{2d}}{24}} = \chi_0(q)\;.
\end{equation}
In these equations we have used the nome $q$ defined as $q\colonequals e^{2\pi i \tau}$. It will be useful later on to also recall that one-point functions are position independent and easily expressed in terms of the operator's zero-mode $o(\mathcal O) \colonequals \mathcal O_0$
\begin{equation}\label{1ptfnct}
\frac{\partial}{\partial z}\langle \mathcal O(z) \rangle_{\mathbb T^2} = 0\;, \qquad \langle \mathcal O(z) \rangle_{\mathbb T^2} =  \frac{(2\pi i)^{h_{\mathcal O}}}{\chi_0(q)} \text{Str}\ o(\mathcal O) \ q^{L_0-\frac{c_{2d}}{24}}\;.
\end{equation}
If the field is not integer-moded or has odd statistics the result is zero. In particular, for the stress-energy tensor $T$, which transforms as $T(z) = (2\pi i w)^2(T(w) - \frac{c}{24 w^2})$, one finds
\begin{equation}\label{stressTensoronept}
\langle T(z) \rangle_{\mathbb T^2} = 2\pi i\ \partial_\tau \log \chi_0(q)\;.
\end{equation}

Let us introduce some notation for the quantities of our most direct interest, namely the the torus two-, and three-point functions
\begin{equation}\label{toruscorr}
b_{IJ}(z_1,z_2;\tau) \colonequals \langle \mathcal O_I(z_1)\ \mathcal O_J(z_2)\rangle_{\mathbb T^2}\;,\qquad c_{IJK}(z_i; \tau) \colonequals \langle \mathcal O_I(z_1)\ \mathcal O_J(z_2)\ \mathcal O_K(z_3) \rangle_{\mathbb T^2}\;.
\end{equation}
Note that $b_{IJ}(z_1,z_2;\tau) = c_{IJ\mathbbm 1}(z_1,z_2,z; \tau)$. We also use the notation $a_I(\tau) = b_{I\mathbbm 1}(z_1,z_2;\tau)$ for the one-point functions.

\subsection{High temperature limit and deformation quantization}
Implementing the intuition outlined above, we would like to take the high-temperature limit $\beta\colonequals -i \tau\rightarrow +0$ of the torus correlation functions of \eqref{toruscorr}. To ensure that only correlators of Higgs branch chiral ring operators survive, we accompany this limit with a specific $\beta$-dependent rescaling of all Schur operators. To find the correct rescaling, we argue as follows. Let us recall that the space of states of the vertex operator algebra or equivalently the vector space of all Schur operators of the four-dimensional superconformal field theory is triply graded by (two-dimensional) conformal weight $h$, $SU(2)_R$ Cartan charge $R$, and $U(1)_r$ charge $r$: $\mathcal V = \bigoplus_{h,R,r} \mathcal V_{h,R,r}$. Note however that the operator product expansions of the associated fields do not respect the $R$-grading, but they do respect an $R$-filtration $\mathcal F_{h,R,r} = \bigoplus_{k\geq 0} \mathcal V_{h,R-k,r}$. See \cite{Beem:2017ooy} for more details. Higgs branch chiral ring operators reside in $\mathcal V_{\text{H}}=\bigoplus_{R}\mathcal V_{h=R,R,r=0}$. It is clear then that a measure for the deviation of a Schur operator from being in the Higgs branch chiral ring is $h-R$. We should additionally take into account that the four-dimensional conformal dimension of a Higgs branch chiral ring operator, $E=2R$, is one unit of $SU(2)_R$ charge larger than that of a three-dimensional one $\widetilde E = R$.\footnote{Here we use that upon dimensional reduction the non-abelian $R$-symmetry can be identified between the four-dimensional UV SCFT and the three-dimensional IR theory.} In total we thus put forward the rescaling of Schur operators to be
\begin{equation}
\mathcal O \rightarrow \beta^{(h-R)+R}\, \mathcal O = \beta^{h}\,\mathcal O\;.
\end{equation}
Quite elegantly, this rescaling can be phrased in terms of the conformal weight of the fields of the VOA, which are known as soon as the vertex operator algebra itself has been identified.

Applying this rescaling to one-point functions, we define
\begin{equation}
 \mathsf a_{I} \colonequals \lim_{\beta\rightarrow +0} \beta^{h_I}\, a_{I}(\tau)\;,
\end{equation}
and trivially find from \eqref{1ptfnct}
\begin{equation}\label{1ptbeta}
\mathsf a_{I} = \lim_{\beta\rightarrow +0}\frac{(2\pi i \beta)^{h_{I}}}{\chi_0(q)} \text{Str}\ o(\mathcal O_I) \ q^{L_0-\frac{c_{2d}}{24}}
\end{equation}

Next, we define, for $\re(z_1)> \re(z_2)$,
\begin{equation}\label{2ptbij}
\mathsf b_{IJ} \colonequals \lim_{\beta\rightarrow +0}\beta^{h_I+h_J}\, b_{IJ}(z_1,z_2;\tau)\;,
\end{equation}
which constitutes a matrix of real numbers. Using Zhu's recursion relations \cite{zhu1996modular}, we can indeed prove explicitly that the position dependence drops out in our limit:
\begin{equation}
\langle \mathcal{O}(z) \mathcal{O}'(z')\rangle_{\mathbb T^2} = \frac{(2\pi i)^{h_{\mathcal O} + h_{\mathcal O'}}}{\chi_0(\tau)}\ \text{Str}\ o(\mathcal O)\, o(\mathcal O')\, q^{L_0 - \frac{c}{24}}  + \sum_{m > 0} \mathcal P_{m+1}(z-z'|\tau) \langle \{\mathcal{O}\mathcal{O}'\}_{m+1}(0)\rangle_{\mathbb{T}^2}\;,
\end{equation}
where
\begin{equation}
\mathcal P_1(z | \tau) \colonequals - \pi i - \partial_z \ln \vartheta_1(z|\tau)\, \qquad \mathcal P_{m+1}(z | \tau) \colonequals - \frac{1}{m!} \partial^{m+1}_z \ln \vartheta_1(z|\tau)\;, \qquad m \ge 1
\end{equation}
and we used the standard notation for the various terms in the operator product expansion
\begin{equation}
\mathcal{O}(z) \mathcal{O}'(z') = \sum_{n} \frac{ \{\mathcal{O}\mathcal{O}' \}_n (z')}{(z-z')^n}\;.
\end{equation}
Using the high-temperature behavior (\ref{asymptoticOfP}) of the functions $\mathcal P_{m+1}$ and of normalized one-point functions $\langle \mathcal{O}\rangle\sim \beta^{- h_\mathcal{O}}$, it is now easy to convince oneself that the second term does not contribute in the limit \eqref{2ptbij}. We thus find
\begin{equation}
\label{2ptbij2}
\mathsf b_{IJ} = \lim_{\beta\rightarrow +0} \frac{(2\pi i \beta)^{h_{I} + h_{J}}}{\chi_0(\tau)}\ \text{Str}\ o(\mathcal O_I)\, o(\mathcal O_J)\, q^{L_0 - c/24}\;.
\end{equation}
Note that this equality for the high-temperature limit of two-point functions is in some sense a trivial extension of \eqref{1ptbeta}. Finally, we define for $\re(z_1)> \re(z_2) > \re(z_3)$
\begin{equation}
{\mathsf  c}_{IJK} \colonequals \lim_{\beta\rightarrow +0} \beta^{h_I+h_J+h_J}\, c_{IJK}(z_1,z_2,z_3;\tau)\;,
\end{equation}
which we expect to be explicitly calculable as\footnote{We have not attempted to prove this statement from first principles, for example again using Zhu's recursion relations.}
\begin{equation}\label{3ptcijk}
{\mathsf  c}_{IJK} = \lim_{\beta\rightarrow +0} \frac{(2\pi i \beta)^{h_{I} + h_{J}+ h_{K}}}{\chi_0(\tau)}\ \text{Str}\ o(\mathcal O_I)\, o(\mathcal O_J)\, o(\mathcal O_K)\, q^{L_0 - c/24}\;.
\end{equation}
The computation of the limit \eqref{3ptcijk} defining the coefficients ${\mathsf  c}_{IJK}$ (and its special cases involving one or two identity operators \eqref{2ptbij} and \eqref{1ptbeta}) can be further simplified. Let us start by considering the vacuum character $\chi_0(\tau)$. Under an S-transformation $\tau \rightarrow \tilde \tau = -1/\tau$ it can be written as
\begin{equation}
\chi_0(\tau) = \sum_j S_{0j} \, \chi_j(\tilde\tau)\;,
\end{equation}
where the sum runs over the elements of the vector-valued modular form of weight zero under $PSL(2,\mathbb Z)$ or $\Gamma^0(2)$ in which the vacuum character of any vertex operator algebra that occurs as the image of the SCFT/VOA map $\mathbb V$ transforms.\footnote{\label{MDECardy}It was shown in \cite{Beem:2017ooy}, as a consequence of the conjecture that the Higgs branch of vacua can be recovered as the associated variety of the vertex operator algebra, that the vacuum character of the vertex operator algebras appearing in the context of the SCFT/VOA correspondence satisfies a modular differential equation. As a result, under modular transformation, the vacuum character is mapped to a linear combination of solutions of that modular differential equation or, in the $\Gamma^0(2)$-modular case, the conjugate one.} The characters $\chi_j(\tilde\tau)$ start off as $\chi_j(\tilde\tau) = e^{2\pi i \tilde\tau (-c/24 + h_j)}(1+\ldots)$, but possibly also involve logarithmic terms. The coefficients $S_{0j}$ are rational numbers. The high-temperature limit of the vacuum character is then easily derived 
\begin{equation}\label{Cardy}
\log\chi_0(\tau) \xrightarrow{\beta \to +0} \frac{\pi i (c- 24h_{\text{min}})}{12\tau}\;,
\end{equation}
where $h_{\text{min}} \colonequals \text{min}_i \, h_i$. For VOAs associated with four-dimensional SCFTs, this Cardy behavior can be related to four-dimensional Weyl anomaly coefficients as \cite{DiPietro:2014bca,Buican:2015ina,Ardehali:2015bla,Beem:2017ooy,Chang:2019uag}\footnote{Here and in the rest of the paper, we assume that $c_\text{eff} > 0$ to avoid divergent high-temperature behavior.}\textsuperscript{,}\footnote{\label{HBdim}Note that if one finds only free hypermultiplets in a generic point of the Higgs branch $\mathcal M_H$ of the SCFT, then $24 (c_{4d}-a_{4d}) = \dim_{\mathbb H}\mathcal M_H$.}
\begin{equation}\label{ceff}
c_{\text{eff}} \colonequals c- 24h_{\text{min}} = 48 (c_{4d}-a_{4d})\;.
\end{equation}
Let us denote the space of states of conformal weight $h_{\text{min}}$ as $M_{\text{min}}$. Assuming that $M_{\text{min}}$ is finite-dimensional,\footnote{One could speculate that the modules of interest here belong to category $\mathcal O$, and thus, upon fully grading, possess finite-dimensional weight spaces. Perhaps this property can be brought to bear to deal with the case of infinite-dimensional, not fully graded spaces  $M_{\text{min}}$.} spanned by Grassman-even states of integer spin, we can straightforwardly write
\begin{equation}\label{cijkZhu}
{\mathsf  c}_{IJK} = (2\pi i)^{h_I+h_J+h_K}\frac{{\text{tr}}_{M_{\text{min}}}o(\mathcal O_I)\, o(\mathcal O_J)\, o(\mathcal O_K)}{{\text{tr}}_{M_{\text{min}}} \mathbbm 1}\;.
\end{equation}
The generalization to more general state-content of $M_{\text{min}}$ requires more care in the trace, as the periodicity properties along the two cycles of the torus are exchanged by the S-transformation. (See also footnote \ref{footnoteperiodicity}.) It is worth noting that expression \eqref{cijkZhu} makes contact with Zhu's non-commutative algebra \cite{zhu1996modular}. Indeed, the product of zero-modes acting on a highest weight state $|\psi\rangle$ can be re-expressed as $o(a)o(b)|\psi\rangle = o(a * b)|\psi\rangle$, where $*$ is Zhu's non-commutative product, turning the argument of the trace in the numerator into $o(\mathcal O_I\, *\, \mathcal O_J\,*\, \mathcal O_K)$.\footnote{This product is defined as 
\begin{equation}
a * b = \oint dz \ a(z) \frac{(1+z)^{h_a}}{z} b(0)|\Omega \rangle \;,
\end{equation}
for two states $a,b$. It is associative, but non-commutative. Zhu's algebra additionally involves a quotient with respect to a certain ideal.
} We leave a more detailed study of this structure for future work.

Having computed the one-, two-, and three-point coefficients ${\mathsf  a}_{I},{\mathsf  b}_{IJ}$, and ${\mathsf  c}_{IJK}$, our next task is to find a new basis of operators $\hat{\mathcal O}_I$ that diagonalizes $\mathsf b_{IJ}$. This Gram-Schmidt process is a recursive task that can be easily performed. Note that the new basis still respects the $R$-filtration, \ie{}, the new operator can only be different from the original one by operators of lower $R$-charges. In particular the diagonalization ensures that in the new basis the one-point functions $\hat{\mathsf a}_I = \hat{\mathsf b}_{I\mathbbm 1} = 0$, for $I\neq \mathbbm 1$. In fact, even stronger, we claim that in the new diagonal basis
\begin{equation}
\hat{\mathsf b}_{II} = 0 \qquad \text{if}\qquad \text{$\mathcal O_I \notin \mathcal V_{\text{H}}$ }\;.
\end{equation}

Implementing the change of basis on the constants ${\mathsf  c}_{IJK}$, which we denote in the new basis as $\hat{{\mathsf  c}}_{IJK}$, we further claim that
\begin{equation}
\hat{\mathsf c}_{IJK} = 0 \qquad \text{if}\qquad \text{$\mathcal O_I \notin \mathcal V_{\text{H}}$ or $\mathcal O_J \notin \mathcal V_{\text{H}}$ or $\mathcal O_K \notin \mathcal V_{\text{H}}$}\;.
\end{equation}
We do not currently know how to prove these claims in all generality. For Lagrangian theories, however, it is easy to convince oneself that they are true. See the next section for more details. Together, the inverse of the (non-zero part of the diagonal) matrix $\hat{\mathsf b}_{IJ}$ and the collection of numbers $\hat{\mathsf c}_{IJK}$ allow one to define
\begin{equation}
\hat{\mathsf c}_{IJ}^{\phantom{IJ}K} \colonequals \sum_L (\hat{\mathsf b}^{-1})^{KL} \, \hat{\mathsf c}_{IJL}\;.
\end{equation}
Here the indices can be understood to only range over $\mathcal V_H$. Other operators have zero three-point constants $\hat{\mathsf c}_{IJL}$ anyway. At last, these define the desired algebra
\begin{equation}\label{starprod}
\hat{\mathcal O}_I \star \hat{\mathcal O}_J = \sum_K \hat{\mathsf c}_{IJ}^{\phantom{IJ}K}\ \hat{\mathcal O}_K\;.
\end{equation}

The algebra defined by \eqref{starprod} is a $\mathbb C^*$-equivariant deformation quantization. These properties immediately follow from the construction presented so far. In particular the $R$-filtration guarantees $\mathbb C^*$-equivariance, and the antisymmetric part of the leading term is the Poisson bracket as follows from an easy computation:
\begin{equation}
[o(a),o(b)] = \sum_{p = 0}^{+\infty} \Big( \ \begin{matrix}
	h_a - 1\\p
\end{matrix} \ \Big) (a_{ - h_a + p + 1}b)_0,
\end{equation}
whose leading $p=0$ term gives $a_{-h_a+1}b = \{a,b\}_{\text{PB}}$ if $a,b\in \mathcal V_H$ \cite{Beem:2017ooy}. The other properties we would like this deformation quantization to possess are less obvious. For Lagrangian theories, they follow indirectly from the arguments presented in the next section. We will assume that they all hold also for non-Lagrangian cases, but it would clearly be desirable to prove this statement. 

Our next task is to answer the question if this algebra is some deformation quantization or precisely the one that follows from the SCFT/Deformation quantization correspondence. Note that even if the algebra satisfies all five properties listed in subsection \ref{subsecdefquant}, and therefore all gauge freedom is (conjecturally) fixed, it still has finitely many free parameters in its period, and it is not \textit{a priori} clear that the procedure outlined in this section selects the superconformal values. In fact, as remarked above and as we will see explicitly in examples below, this is not the case if the Coulomb branch $R$-symmetry of the three-dimensional IR theory is a mixture of the UV $U(1)_r$ symmetry with accidental (topological) symmetries. This situation arises whenever the $U(1)_r$ charges of the four-dimensional UV theory are not quantized in half-integer units. Given the discussion at the end of subsection \ref{subsecdefquant} stating that Fayet-Iliopoulos parameters (modded out by the Weyl group of Coulomb branch flavor symmetries) are in one-to-one correspondence with the period of the quantization, it comes as no surprise that having to turn on imaginary Fayet-Iliopoulos parameters directly affects the period.

%% file: sections/tests.tex
\section{Lagrangian proof of proposal}\label{tests}

Our proposal to extract from the vertex operator algebra associated with a four-dimensional $\mathcal N=2$ SCFT the deformation quantization of the Higgs branch of vacua of its dimensional reduction is in essence based on taking the high-temperature limit of torus correlation functions of the VOA. Computing torus correlators is a difficult task in general, although tools like Zhu's recursion relations are often helpful \cite{zhu1996modular}. There is, however, a subset of VOAs whose torus correlation functions can be computed relatively easily, namely those associated with four-dimensional Lagrangian theories. As mentioned before, supersymmetric localization techniques applied to the four-dimensional theory placed on $S^3 \times_q S^1$ lead one to a computational recipe of torus correlators in terms of explicit contour integrals \cite{Pan:2019bor,Dedushenko:2019yiw}. Their integrands have a transparent structure that allows one to extract the deformation quantization of the dimensionally reduced theory. In particular, in this section we will show in detail that the high-temperature limit of these torus correlation functions directly reduces to twisted Higgs branch correlation functions of the three-dimensional theory on the three-sphere. From these one can straightforwardly deduce the desired deformation quantization \cite{Dedushenko:2016jxl}. In fact, a detailed understanding of the Lagrangian procedure is what allowed us to formulate our general proposal. 

\subsection{Lagrangian proof}\label{lagrtest}
Recall from \eqref{vacchar=index} that the unflavored Schur index $I_S^{(\mathcal T_{4d})}(q)$ of any four-dimensional $\mathcal{N} = 2$ superconformal field theory $\mathcal{T}_\text{4d}$ equals the vacuum character of the associated vertex operator algebra $\mathbb{V}(\mathcal{T}_\text{4d})$. Standard arguments further identify the Schur index with the partition function of $\mathcal T_{4d}$ placed supersymmetrically on an $S^3 \times_q S^1$ background, see \eqref{index=partfnct}.\footnote{As explained in detail in \cite{Pan:2019bor}, the supersymmetric background is described by the metric $ds^2 = \ell^2 \cos^2\theta (d\varphi - i \beta_+ dt)^2 + \ell^2 \sin^2 \theta (d\chi - i \beta_- dt)^2 + \ell^2 d\theta^2 + (-i\ell\tau + \ell \beta_+)^2 dt^2$, for some constants $\beta_\pm$ and $\tau$ such that $\re \tau + i \beta_+ = 0$. To ensure the background preserves supersymmetry one should also turn on an $SU(2)_\mathcal{R}$ and $U(1)_r$ background gauge field. The locus $\theta = 0$ defines a torus $\mathbb{T}^2 \subset S^3 \times_q S^1$. On this torus one can insert (almost-)BPS operators: the curved-space counterpart of the twisted translated Schur operators on flat space. Correlation functions of these operators admit a localization computation \cite{Pan:2019bor}, leading to the integral formula we present in the main text.} Concretely, for an $\mathcal{N} = 2$ superconformal gauge theory with gauge algebra $\mathfrak{g}$ and hypermultiplets transforming in the representation $\mathcal{R}$ of $\mathfrak{g}$, the Schur index $I_S$/vacuum character of the associated VOA $\chi_0$/partition function on $S^3 \times_q S^1$ can be written as a contour integral\footnote{See appendix \ref{specialfuctions} for details on Jacobi theta and Dedekind eta functions.}
\begin{align}\label{contourindex}
\chi_0(q) &= \oint \prod_{A = 1}^{\rank \mathfrak{g}} \frac{da_A}{2\pi i a_A} Z_\text{one-loop}(a; q) \nonumber\\
&=\oint \prod_{A = 1}^{\rank \mathfrak{g}} \frac{da_A}{2\pi i a_A} \frac{(-i)^{\rank \mathfrak{g} - \dim \mathfrak{g}}}{|W|}  \  \eta(\tau)^{3\rank \mathfrak{g} -\dim \mathfrak{g}} \prod_{\alpha \in \Delta} \vartheta_1(\alpha(\mathfrak{a})|\tau)  \prod_{\rho \in \mathcal{R}} \frac{\eta(\tau)}{\vartheta_4(\rho(\mathfrak{a})|\tau)} \; .
\end{align}
Here the gauge fugacity $a$ takes values in the Cartan torus, \ie{}, $a = e^{2\pi i \mathfrak{a}}$ with $\mathfrak{a}$ in the Cartan subalgebra of $\mathfrak{g}$. Furthermore, $\Delta$ denotes the (nonzero) roots of $\mathfrak{g}$ and $\prod_{\rho \in \mathcal{R}}$ denotes a product over all weights in the representation $\mathcal{R}$. Finally, $|W|$ is the order of the Weyl group of $\mathfrak g$.

Torus correlation functions of vertex operators of $\mathbb{V}[\mathcal{T}_\text{4d}]$ are identified with those of the corresponding twisted-translated Schur operators, \ie{}, a representative of the cohomology class defining the vertex operator in question, inserted on a torus $\mathbb{T}^2 \subset S^3 \times_q S^1$. Localization techniques provide an explicit contour integral expression for these correlation functions \cite{Pan:2019bor},
\begin{equation}\label{corrlagr4d}
  \Big\langle \prod_i \mathcal{O}_i(z_i)\Big\rangle_{\mathbb{T}^2}
  = \frac{1}{\chi_0(q)} \oint \prod_{A = 1}^{\rank \mathfrak{g}} \frac{da_A}{2\pi i a_A} Z_\text{one-loop}(a; q) \, \big\langle \prod_i \mathcal{O}_i(z_i) \big\rangle_\text{GT} \;.
\end{equation}
Here the coordinates $z = \varphi + \tau t$ parameterize the torus $\mathbb{T}^2$ with complex modulus $\tau$; they are doubly-periodic with periods $1$ and $\tau$. We require the $z_i$'s to have distinct $\varphi_i = \re z_i$. Furthermore, $\langle \mathcal{O} \rangle_\text{GT}$ is the correlation function of the vertex operators in a Gaussian theory defined in terms of a $bc \beta \gamma$ system. In more detail, the corresponding twisted-translated Schur operators are gauge-invariant composites of the vector multiplet gaugini $\lambda, \tilde \lambda$, hypermultiplet scalars $Q_A$, $A = 1,2$, and the holomorphic covariant derivative $D_z$. Note that the letters  $\lambda, \tilde \lambda$ and $Q_A$ transform in the adjoint representation and the matter representation $\mathcal{R}$ of the gauge algebra $\mathfrak{g}$ respectively. Using Wick's theorem, the Gaussian theory is fully defined by specifying the propagators of these letters:
\begin{align}
 G^{\mathbb{T}^2}_{AB}(z, \bar z){^{\rho'}_\rho} &\colonequals \langle Q_A^{\rho'}(z)Q_{B\rho}(0)\rangle_\text{GT}
  = \ - \frac{i \epsilon_{AB}}{(2\pi)^3 \ell^2} e^{- 2\pi i \rho(\mathfrak{a}) \frac{\bar z - z}{\bar \tau - \tau}}\frac{\vartheta_1'(0|\tau)}{\vartheta_4(\rho(\mathfrak{a})|\tau)} \frac{\vartheta_4(z + \rho(\mathfrak{a})|\tau)}{\vartheta_1(z|\tau)} \delta^{\rho'}_\rho \ , \label{GP1}\\
G^{\mathbb{T}^2}(z, \bar z)_{ww'}&\colonequals  \langle \tilde \lambda_{w'}(0)\lambda(z)_w\rangle_\text{GT} =  \ - \frac{1}{\ell^2} \eta(\tau)^3 
    e^{2\pi i w(\mathfrak{a}) \frac{z - \bar z}{\bar \tau - \tau}}
    \partial_z \left[
      \frac{\vartheta_1(w(\mathfrak{a}) + z|\tau)}{\vartheta_1(z|\tau) \vartheta_1(w(\mathfrak{a})|\tau)}
    \right]\delta_{w w'} \ .\label{GP2}
\end{align}
As before, $\mathfrak{a}$ is defined by $a = e^{2\pi i \mathfrak{a}}$, $\rho, \rho'$ are weights of $\mathcal{R}$, while $w, w'$ are weights of the adjoint representation $\text{Adj}_\mathfrak{g}$.

With these ingredients one can compute any torus correlation function of vertex operator algebras associated with Lagrangian SCFTs. According to our proposal, we should consider their high-temperature limit. Let us thus take $q = e^{-2\pi \beta}$ and $a = e^{ - 2\pi \beta \sigma}$. Then $\beta$  parameterizes the radius of the temporal circle of $S^3 \times_q S^1$. Sending $\beta \to +0$ while keeping $\sigma$ fixed shrinks that circle and effectively dimensionally reduces $\mathcal{T}_\text{4d}$ to a three-dimensional theory $\mathcal{T}_\text{3d}$ on the three-sphere. Indeed, one easily computes the high-temperature behavior of the contour integral in \eqref{contourindex}:
\begin{equation}
\chi_0(q) \xrightarrow{\beta \to +0} e^{\frac{\pi}{6\beta} (\dim \mathcal{R} - \dim \mathfrak{g})} \int \prod_{A = 1}^{\rank \mathfrak{g}} d\sigma_A \frac{\prod_{\alpha \in \Delta} 2\sinh \pi \alpha(\sigma)}{\prod_{\rho \in \mathcal{R}}2\cosh \pi \rho(\sigma)}\;,
\end{equation}
where we recognize the prefactor as capturing the Cardy behavior \eqref{Cardy}. Here, $\frac{\pi}{6\beta} (\dim \mathcal{R} - \dim \mathfrak{g}) = 4\pi  \frac{c_{4d} - a_{4d}}{\beta}$ which matches the expected effective central charge in \eqref{ceff}.\footnote{Recall that for a Lagrangian theory with $n_h$ hypermultiplets and $n_v$ vector multiplets one has $a_{4d} = \frac{5 n_v + n_h}{24}$ and $c_{4d}=\frac{2 n_v + n_h}{12}$. Hence $c_{4d}-a_{4d} = \frac{n_h-n_v}{24}$. } The leftover integral is precisely the $S^3$ partition function $Z^{S^3}_{\mathcal{T}_\text{3d}}$ of a three-dimensional $\mathcal N=4$ supersymmetric theory $\mathcal{T}_\text{3d}$ with UV Lagrangian description in terms of a gauge theory with hypermultiplets transforming in representation $\mathcal R$ of the gauge group $G$ with Lie algebra $\mathfrak{g}$. More generally, we can take the high-temperature limit of any correlation function computed as in \eqref{corrlagr4d}.\footnote{Note that the numerator and denominator in \eqref{corrlagr4d} share the same Cardy behavior which thus cancels in the high temperature limit.} To do so, it clearly suffices to analyze the high-temperature behavior of the Gaussian propagators \eqref{GP1} and \eqref{GP2}, and their spatial derivatives.

We proceed in two steps: first we show that the high-temperature limit of correlation functions of Higgs branch chiral ring operators results in the desired deformation quantization. Next we argue that correlation functions involving any other type of operator ultimately do not play a role in the quantization, as expected. Let us thus consider the Higgs branch chiral ring operators, \ie{}, composites of the hypermultiplet scalars $Q_A$ without any derivatives. When $ \frac{1}{2}> \re{}z = \varphi > 0$,\footnote{The result is similar for $ - \frac{1}{2} < \varphi < 0$}
\begin{align}\label{reductionprop}
  G^{\mathbb{T}^2}_{AB}(z, \bar z)_\rho^{\rho'} & \ \xrightarrow{\beta \to +0} - \frac{i \epsilon_{AB}}{(2\pi \ell)^2 \ell \beta}\Big[\operatorname{sign}(\varphi) + \tanh \pi \rho(\sigma)\Big]e^{-\rho(\sigma) \varphi} \delta_\rho^{\rho'} \equiv \frac{1}{\beta} G^{\mathbb{S}^1}_{AB}(\varphi; \sigma)_\rho^{\rho'}\ ,
\end{align}
where $G_{AB}^{\mathbb{S}^1}(\varphi, \sigma)_\rho^{\rho'}$ is the twisted Higgs branch propagator of \cite{Dedushenko:2016jxl}. Note that the high-temperature limit of $G_{AB}^{\mathbb{T}^2}(z, \bar z)$ exhibits the singularity $\beta^{-1} = \beta^{- (h_{\tilde Q} + h_Q)}$. Hence, upon rescaling by $\beta^{\sum h}$, the torus correlation functions of Higgs branch operators have nontrivial high-temperature limit, and precisely equal those computed in \cite{Dedushenko:2016jxl}. Let us give a few more details. The authors of \cite{Dedushenko:2016jxl} performed a localization computation of $\mathcal N=4$ supersymmetric theories on the three-sphere $S^3$ that allows additional insertions of twisted-translated Higgs branch chiral ring operators on a circle $S^1\subset S^3$. Their result mirrors the expression \eqref{corrlagr4d} in the obvious manner
\begin{equation}\label{corrlagr3d}
\Big\langle \prod_i \widetilde{ \mathcal{O}}_i(\varphi_i)\Big\rangle_{S^1}
  = \frac{1}{Z^{S^3}_{\mathcal{T}_\text{3d}}} \int \prod_{A = 1}^{\rank \mathfrak{g}} d\sigma_A \frac{\prod_{\alpha \in \Delta} 2\sinh \pi \alpha(\sigma)}{\prod_{\rho \in \mathcal{R}}2\cosh \pi \rho(\sigma)} \, \big\langle \prod_i \widetilde {\mathcal{O}}_i(\varphi_i) \big\rangle_\text{GT} \;,
\end{equation}
and we have just confirmed that we recover the propagator defining this effective one-dimensional Gaussian theory. Moreover, as was shown in \cite{Dedushenko:2016jxl}, the resulting two- and three-point functions precisely capture the deformation quantization satisfying all five properties that we are aiming to recover after performing a Gram-Schmidt process (if necessary) to diagonalize the two-point functions. 

Next we consider correlators $\langle \prod_i \mathcal{O}_i(z_i)\rangle_\text{GT}$ involving non-Higgs branch chiral ring operators, \ie{}, composites containing derivatives of $Q, \widetilde Q$ and/or involving gaugini. We will refer to these ingredients as non-Higgs branch letters; naturally, $Q, \widetilde Q$ themselves will be called Higgs branch letters. To show that these correlators ultimately do not contribute to the deformation quantization, we analyze the high-temperature limit of all possible cross-contractions, \ie{}, contractions between letters belonging to different operators, and self-contractions, \ie{}, contractions among letters belonging to the same operator. Cross-contractions lead to the quantities $D_z^n G(z, \bar z)$ or $D_z ^n G_{AB}(z, \bar z)$ with $\varphi = \re{}z \ne 0$ and $n\geq 0$, while self-contractions involve these quantities for $\varphi = 0$.

The high-temperature behavior of the gaugino propagator can be easily obtained,
\begin{equation}
 G^{\mathbb{T}^2}(z, \bar z)_{ww'}   \ \xrightarrow{\beta \to +0} - \frac{1}{2\ell^2 \beta} \Big[ \coth\pi \sigma + \operatorname{sign}(\varphi) \Big] e^{- w(\sigma)\varphi} \delta_{ww'} \equiv \frac{1}{\beta} G^{\mathbb{S}^1}(\varphi; \sigma)_{ww'}\ .
\end{equation}
It is crucial to note that it exhibits a $\beta^{-1}$ singularity, milder than $\beta^{- (h_\lambda + h_{\tilde \lambda})}$ by a factor of $\beta$.
Moreover, as long as $\varphi \ne 0$, we also find $\beta^{-1}$ singularities in the presence of derivatives,
\begin{align}
  D_z^n G^{\mathbb{T}^2}_{AB}(z, \bar z)_\rho^{\rho'} \xrightarrow{\beta \to +0} \frac{1}{\beta} D_\varphi^n G^{\mathbb{S}}_{AB}(\varphi)_\rho^{\rho'},
  \qquad
  D_z^n G^{\mathbb{T}^2}(z, \bar z)_{ww'} \xrightarrow{\beta \to +0} \frac{1}{\beta} D_\varphi^n G^{\mathbb{S}}(\varphi)_{ww'} \ .
\end{align}
Thus, upon rescaling with the appropriate $\beta^{h}$ factors and taking the limit $\beta\rightarrow +0$, cross-contractions of non-Higgs branch letters are all suppressed.

Subtleties arise when we consider self-contractions of non-Higgs branch letters. Indeed, when $\re z = 0$, one can verify that
\begin{align}
  D_z^{n > 0}G_{AB}^{\mathbb{T}^2}(z, \bar z)_\rho^{\rho'} \xrightarrow{\beta \to +0} & \ - \frac{i\epsilon_{AB}}{\beta^{n + 1}(2\pi)^3 \ell^2} \pi^2  \bigg(-i\frac{\partial}{\partial t}\bigg)^{n-1} \left[\frac{1}{\sin^2 \pi t}\right] \delta_\rho^{\rho'}\epsilon_{AB} \ ,\\
  D_z^{n}G^{\mathbb{T}^2}(z, \bar z)_{ww'} \xrightarrow{\beta \to +0} & \ - \frac{\epsilon_{AB}}{\beta^{n + 2}(2\pi) \ell^2} \pi^2  \bigg(-i\frac{\partial}{\partial t}\bigg)^n \left[\frac{1}{\sin^2 \pi t}\right] \delta_{ww'} \ ,
\end{align}
which contain the singularities
\begin{align}
  \beta^{-(n + 1)} = \beta^{- (h_Q + h_Q + h_{\partial^n})}\ , \qquad
  \beta^{- (n + 2)} = \beta^{- (h_\lambda + h_{\tilde \lambda} + h_{\partial^n})} \ .
\end{align}
Upon introducing the appropriate factors $\beta^h$, the singularities in $\beta$ cancel. The $t$-independent pieces of the above expressions are used in the standard point-splitting procedure. Hence, we find a nontrivial high-temperature limit when performing self-contractions.

After performing all possible Wick contractions, each term in the computation of $\langle \prod_i \mathcal{O}_i(z_i)\rangle_\text{GT}$ can be factorized into three types of factors: factors arising from i) self-contractions of non-Higgs branch letters, ii) cross/self-contractions solely between Higgs branch letters, and iii) cross-contractions involving non-Higgs branch letters. The former two types of contractions contain precisely the necessary singularities in $\beta$ to ensure their own survival in the high-temperature limit, while the third type does not. Consequently, terms that contain non-Higgs branch cross-contractions vanish in the limit. In other words, in the computation of the correlator $\langle \prod_i \mathcal{O}_i(z_i)\rangle_\text{GT}$ only those terms containing solely the first two types of factors actually survive the limit, and in particular, every non-Higgs branch letter must engage in some self-contraction to ensure that. 

To continue the argument, we bring to bear the Gramm-Schmidt process, which instructs us to diagonalize all two-point functions. In particular, for any non-Higgs branch operators $\mathcal O_{\text{non-HB}}$, this means that we should subtract all possible maximal self-contractions involving non-Higgs branch letters,
\begin{align}
\widehat{\mathcal{O}}_{\text{non-HB}} \equiv \mathcal{O}_{\text{non-HB}} - \text{maximal non-Higgs branch self-contractions} \ .
\end{align}
Indeed, the two-point function of a non-Higgs branch operator $\widehat{\mathcal{O}}_{\text{non-HB}}$ with a Higgs branch operator $\mathcal{O}_{\text{HB}}$  can then be computed to be trivial,
\begin{align}
  \lim_{\beta \to +0} \beta^{h_{\text{non-HB}} + h_{\text{HB}}}\left\langle \widehat{\mathcal{O}}_{\text{non-HB}}(z) \mathcal{O}_{\text{HB}}(w)\right\rangle
  = 0 \ ,
\end{align}
since all the non-Higgs branch letters in $\mathcal{O}_{\text{non-HB}}$ contribute only via self-contractions which are already subtracted away in $\widehat{\mathcal{O}}_{\text{non-HB}}$. The same goes for two-point functions between two non-Higgs branch operators. The net result is then the decoupling of non-Higgs branch operators in two-point functions, and in fact in more general correlators. This then completes the proof that our proposal works for all Lagrangian four-dimensional theories.

\subsection{Lagrangian examples}
Before moving on to non-Lagrangian tests in the next section, let us quickly summarize a few simple Lagrangian examples where the relevant vertex operator (sub)algebras are current algebras. We perform the general analysis of current algebras $\widehat{\mathfrak{g}}_k$ in appendix \ref{AKM}; the final result of the current-current star product is given by\footnote{We use conventions in which the longest root has length squared $\psi^2=2$.}
\begin{equation}\label{currentstar}
  {J}^a \star {J}^b = ({J}^a {J}^b) + \zeta i f^{ab}_{\phantom{ab}c}  J^c + \zeta^2 \left(\frac{k}{3} - \frac{ (k + h^\vee)c_\text{eff}}{3 \dim \mathfrak{g}}\right) \kappa^{ab} \;,
\end{equation}
where $f^{ab}_{\phantom{ab}c}$ denotes the structure constants and $\kappa^{ab}$ the Killing form. Furthermore, we used the effective central charge defined in \eqref{ceff}, $h^\vee$ is the dual coxeter number of $\mathfrak g$, and $\dim \mathfrak{g}$ denotes its dimension. Let us introduce a convenient parameter capturing the $\zeta^2$ coefficient:
\begin{equation}\label{mudef}
\mu = \frac{3}{8}\left(\frac{k}{3} - \frac{ (k + h^\vee)c_\text{eff}}{3 \dim \mathfrak{g}}\right)\;.
\end{equation}

For $\widehat{\mathfrak{su}}(2)_k$ current algebras, it is customary to reorganize the currents as
\begin{equation}\label{XYZbasis}
  X = \frac{i}{2\sqrt{2}} (J^1 + i J^2), \qquad
  Y = \frac{i}{2\sqrt{2}} (J^1 - i J^2), \qquad
  Z = \frac{1}{2\sqrt{2}} J^3 \ ,
\end{equation}
leading to the star products
\begin{align}
  Z \star Z = & \ (ZZ) \qquad\quad  + \frac{1}{3} \mu \zeta^2 \nonumber \\
  Z \star X = & \ (Z X) + \frac{\zeta}{2} X \label{dffreeHM}\\
  Z \star Y = & \ (Z Y) - \frac{\zeta}{2} Y \nonumber\\
  X \star Y = & \ (ZZ) - \zeta Z \quad - \frac{2}{3}\mu \zeta^2 \nonumber\ .
\end{align}
A first elementary example is $\widehat{\mathfrak{su}}(2)_{-\frac{1}{2}}$, the $\mathbb{Z}_2$-invariant vertex operator subalgebra of the associated VOA of a free four-dimensional hypermultiplet. In this case the vacuum character reads
\begin{align}
\chi_0(q) =\frac{1}{2}\left( \frac{\eta(\tau)}{\vartheta_3(0|\tau)} + \frac{\eta(\tau)}{\vartheta_4(0|\tau)}\right)\;,
\end{align}
whose high-temperature behavior can be easily computed using the formulas in appendix \ref{specialfuctions}, resulting in $c_\text{eff} = 2$ and thus yielding $\mu = - \frac{3}{16}$.\footnote{The value of $c_{\text{eff}}$ could also have been obtained by using footnote \ref{HBdim} in combination with formula \eqref{ceff}, as of course the quaternionic dimension of the relevant Higgs branch is one.} This indeed agrees with the deformation quantization of the $\mathbb{Z}_2$ gauge theory of the free hypermultiplet \cite{Beem:2016cbd}. Note that no diagonalization is required in this example, and that both the $\widehat{\mathfrak{su}}(2)_{-\frac{1}{2}}$ torus correlation functions and the star products can also be analyzed directly using the Lagrangian machinery introduced above.

A second example serves to probe the validity of our results in cases where $c_{\mathrm{eff}}\leq 0$. We focus on the boundary case $c_{\mathrm{eff}}=0$, realized most simply in four-dimensional $\mathcal{N} = 4$ super Yang-Mills theory with gauge group $SU(2)$. Its associated VOA is the small $\mathcal N=4$ superconformal algebra at $c=-9$ \cite{Beem:2013sza}. We can focus on its current subalgebra $\widehat{\mathfrak{su}}(2)_{-\frac{3}{2}}$ generated by Higgs branch chiral ring operators, if we assume that our derivation that all VOA correlators involving fermionic letters (or derivatives) vanish in our high-temperature limit still holds. Then the supercurrents are effectively removed, and similarly the canonical stress-energy tensor ultimately decouples.\footnote{Recall that the stress tensor is cohomologous to the Sugawara stress tensor in the small $\mathcal N=4$ superconformal algebra at $c=-9$. Its status as zero in the deformation quantization thus implies that the Higgs branch relation $\kappa_{ab}J^aJ^b = 0$ indeed holds.} The VOA's character reads
\begin{align}
  \chi_0 = \frac{1}{2} \oint \frac{da}{2\pi i a} \frac{\vartheta_1(2\mathfrak{a}|\tau) \vartheta_1(-2\mathfrak{a}|\tau) \eta(\tau)^3}{\vartheta_4(2\mathfrak{a}|\tau) \vartheta_4(-2\mathfrak{a}|\tau) \vartheta_4(0|\tau)} \ .
\end{align}
Its leading Cardy behavior is indeed trivial, but the subleading behavior is logarithmic, as can be verified explicitly by solving the conjugate modular differential equation (see footnote \ref{MDECardy}). This logarithmic term, however, is not sufficiently singular to survive our $\beta\rightarrow +0$ limit in \eqref{JJ2ptf}-\eqref{JJJ3ptf}. Indeed, $\frac{d}{d\tau} \log\log \tilde q = \frac{d}{d\tau} \log \frac{-2\pi i}{\tau} = -\frac{1}{\tau}$. Computing $\mu$ defined in \eqref{mudef} then results once again in the value $\mu = - \frac{3}{16}$.

The naive dimensional reduction of four-dimensional $\mathcal{N} = 4$ super Yang-Mills theory is three-dimensional $\mathcal{N} = 8$ super Yang-Mills theory with the same gauge group. The $S^3$-partition function then reads
\begin{equation}
  Z = \lim_{\beta \to +0} \frac{1}{|W|}\int_{- \frac{\pi}{\beta}}^{+\frac{\pi}{\beta}} d\sigma
  \frac{2 \sinh(2\pi \sigma) 2\sinh(-2\pi \sigma)}{2\cosh(2\pi \sigma)2 \cosh(-2\pi \sigma) 2\cosh(0)} \ .
\end{equation}
This integral suffers from an obvious $\beta^{-1}$ divergence as $\beta \to 0$, which reflects the subleading logarithmic divergence of the high-temperature limit of the Schur index. The divergent nature of the three-sphere partition function also indicates that the naive dimensional reduction of four-dimensional $\mathcal{N} = 4$ SYM is a bad theory. Nevertheless, keeping the parameter $\beta$ as a regulator, one can easily compute the relevant normalized correlation functions of Higgs branch operators. The end-results of these computation admit a finite $\beta\rightarrow 0$ limit, which naturally agree with the high-temperature behavior of the corresponding Schur correlation functions. However, this procedure does not seem to have a clear, intrinsically three-dimensional interpretation. A more physical approach would be to consider a good three-dimensional $\mathcal N=4$ UV description of the same IR fixed-point instead. For example, up to a decoupled free hyper, an $SU(2)\times U(1)$ gauge theory with bifundamental matter and an adjoint $SU(2)$ hypermultiplet. Computations in this theory, however, do not agree with the high-temperature limit of the Schur correlation functions we have presented above. We leave a careful analysis of this example and other theories with $c_\text{eff}\leq 0$ for future work.

\section{Non-Lagrangian examples}\label{nonlagrtests}
In this section, we turn our attention to non-Lagrangian SCFTs with known VOAs. In particular we analyze a class of examples with trivial Higgs branches to probe the decoupling of non-Higgs branch operators in the non-Lagrangian setting -- we will see that an argument that parallels the one given in the Lagrangian case applies once again --, and study various instances of theories whose associated VOAs are current algebras. We compute a variety of star-products of the resulting deformation quantization and confirm that their period takes different values than their superconformal values if the $R$-symmetries mix with accidental symmetries. 

\subsection{\texorpdfstring{$(A_1, A_{2n})$}{(A1,A2n)} Argyres-Douglas theories}
As a first non-Lagrangian test of our proposal we consider the $(A_1, A_{2n})$ Argyres-Douglas theories. Their associated vertex operator algebras are the Virasoro minimal models $\text{Vir}_{2, 2n + 3}$ \cite{Cordova:2015nma,Beem:2017ooy}. As these theories have a trivial Higgs branch, we should find a trivial deformation quantization. In other words, this example will test if the Virasoro stress tensor is removed properly in our high-temperature limit.

The vertex operator algebra $\text{Vir}_{2, 2n + 3}$ is strongly generated by the stress tensor $T(z)$. Its central charge is $c=-\frac{2n(6n+5)}{2n+3}$. On the torus, the stress tensor always has a non-zero one-point function $t \colonequals \langle T(z)\rangle_{\mathbb{T}^2}$, which can be computed in terms of the derivative of the vacuum character, see \eqref{stressTensoronept}. Combined with the Cardy behavior of \eqref{Cardy}, we find a non-zero high-temperature limit:
\begin{equation}
\lim_{\beta\to +0}\beta^2 t = \frac{(2\pi i)^2 c_\text{eff}}{24}\;,
\end{equation}
and thus a non-diagonal $\mathsf{b}_{IJ}$, as clearly $\mathsf{b}_{T\mathbbm 1}\neq 0$. A Gram-Schmidt procedure is thus needed to diagonalize the matrix $\mathsf{b}_{IJ}$, instructing us in particular to study correlation functions of $\hat{T}(z) \equiv T (z) - t$.

From \eqref{2ptbij2} and \eqref{3ptcijk}, the high-temperature limit of the rescaled stress tensor correlation functions reads ($n = 2, 3$ for our purposes)
\begin{align}
  \lim_{\beta \to +0} \beta^{2n} \big\langle \prod_{i = 1}^n T(z_i) \big\rangle_{\mathbb{T}^2}
  = & \ \lim_{\beta \to +0} \frac{(2\pi i \beta)^{2n}}{\chi_0(\tau)}\operatorname{Str} \left(L_0 - \frac{c}{24}\right) ^n q^{L_0 - \frac{c}{24}} \\
  = & \  \lim_{\beta \to +0} \frac{ (2\pi i)^n \beta^{2n}}{\chi_0(\tau)}\partial_\tau^n  \chi_0(q)\ ,
\end{align}
where we have used that $\partial_\tau = 2\pi i q \partial_q$. Combined with the Cardy behavior of $\chi_0$, we obtain
\begin{equation}
  \lim_{\beta \to +0} \beta^{2n} \big\langle \prod_{i = 1}^n T(z_i) \big\rangle_{\mathbb{T}^2} = \left(\frac{(2\pi i)^2 c_\text{eff}}{24}\right)^n = \lim_{\beta\to +0} \beta^{2n} t^n\;.
\end{equation}
This equality immediately leads to a trivial high-temperature limit of all correlation functions of $\hat{T}$, thanks to the identity \eqref{decomposeTcorrelationfn}. Moreover, the vanishing behavior remains true when including composites of $\hat T$, constructed as the regular part of the coincident limit of products of $\hat T$, or in the presence of spatial derivatives. We conclude that any Virasoro algebra results in a trivial deformation quantization and in particular so does $\text{Vir}_{2, 2n+3}$.

In fact, we can prove the stronger statement that the stress tensor $T$ of any VOA trivializes when considering our high-temperature limit and thus decouples from the resulting deformation quantization. Indeed, let us first write as before
\begin{equation}
  \lim_{\beta\to+0} \big\langle \prod_{j=1}^n \beta^{h_j}\mathcal{O}_j(z_j) \big\rangle_{\mathbb{T}^2} = \lim_{\beta \to +0} \frac{(2\pi i \beta)^{\sum_j h_j}}{\chi_0(\tau)} \text{Str}\prod_j o(\mathcal{O}_j) q^{L_0 - \frac{c}{24}} \;.
\end{equation}
Additionally inserting a collection of stress tensors $\prod_{l=1}^m \beta^2 T(z_l)$ simply produces an $m$-fold derivative with respect to $\tau$ on the supertrace $\operatorname{Str}\prod_i o(\mathcal{O}_i) q^{L_0 - \frac{c}{24}}$ \cite{zhu1996modular,Beem:2017ooy}
\begin{equation}
\lim_{\beta\to+0} \big\langle \prod_{l=1}^m \beta^2 T(z_l) \prod_{j=1}^n \beta^{h_j}\mathcal{O}_j(z_j) \big\rangle_{\mathbb{T}^2} = \lim_{\beta \to +0} \frac{(2\pi i \beta)^{\sum_j h_j}}{\chi_0(\tau)}(2\pi i\beta^{2})^m \partial_\tau^m \left(\text{Str}\prod_j o(\mathcal{O}_j) q^{L_0 - \frac{c}{24}}\right) \;.
\end{equation}
The high-temperature behavior of the latter can be inferred from its properties under S-transformation. A reasoning similar to the one above then allows us to conclude that all $\langle \prod_{j=1}^m \beta^2\hat{T}(z_j) \prod_i \beta^{h_i}\mathcal{O}_i(z_i)\rangle$ vanish in the $\beta\to +0$ limit. In appendix \ref{app:stresstensor} we present an alternative analysis of the decoupling of the stress tensor based on Ward identities.

\subsection{\texorpdfstring{$(A_1, D_{2n+1})$}{(A1,D2n+1)} Argyres-Douglas theories}
The next class of examples we consider are the $(A_1, D_{2n+1})$ Argyres-Douglas SCFTs. Their associated VOAs are given by $\widehat{\mathfrak{su}}(2)_{k = - \frac{4n}{2n+1}}$ current algebra \cite{Cordova:2015nma,Beem:2017ooy}. Their vacuum characters read
\begin{align}
	\chi_0^{(n)}(\mathfrak{a})
	= \frac{\vartheta_1(\mathfrak{a}|(2n + 1)\tau)}{\vartheta_1(\mathfrak{a}|\tau)} \xrightarrow{\mathfrak{a} \to 0}  \frac{\eta( (2n + 1)\tau)^3}{\eta(\tau)^3}\;,
\end{align}
which have Cardy behavior determined in terms of
\begin{equation}
c_{\text{eff}} = \frac{6n}{1+ 2n}\;.
\end{equation}
Using the results of appendix \ref{AKM} on the deformation quantization to which current algebras reduce, and the redefinition of currents as in \eqref{XYZbasis}, we find the elementary star products
\begin{align}
  Z \star Z = & \ (ZZ) \qquad \quad - \frac{n(1+n)}{3(1+2n)^2} \zeta^2 \nonumber \\
  Z \star X = & \ (Z X) + \frac{\zeta}{2} X \nonumber\\
  Z \star Y = & \ (Z Y) - \frac{\zeta}{2} Y \\
  X \star Y = & \ (ZZ) -\zeta Z \quad  + \frac{2n(1+n)}{3(1+2n)^2} \zeta^2 \nonumber\ .
\end{align}

These results can be compared against a three-dimensional computation. Let us start with identifying the three-dimensional $\mathcal{N} = 4$ supersymmetric theory to which the $(A_1, D_{2n+1})$ theory flows. The circle reduction of the vacuum characters $\chi_0^{(n)}$ can be easily performed: 
\begin{equation}
  \lim_{\beta \to +0} e^{ - \frac{n \pi i}{2 + 4n} \frac{1}{\tau}} \chi_0^{(n)}(a, q)
  = \frac{\sin(\frac{\pi}{1+2n})}{2\sqrt{1+2n}} \ \frac{2 \sin \frac{\pi i m}{2n+1}}{\sinh (\pi m) \sinh(\frac{\pi i}{2n+1})} \ .
\end{equation}
where we removed the Cardy behavior and set $\mathfrak{a} = \beta m$. Up to prefactors irrelevant for our current purposes,\footnote{These factors indicate the presence of additional free twisted hypermultiplets, besides the interacting $\mathcal{N} = 4$ SQED with two flavors, in the infrared \cite{Giacomelli,Benvenuti:2018bav}. Computationally, however, they cancel when calculating normalized correlation functions of the $SU(2)$ flavor symmetry moment maps of the interacting part.} this result can be recognized as the $S^3$-partition function of three-dimensional $\mathcal{N} = 4$ SQED with two flavors in the presence of an imaginary FI-parameter (see, \eg{}, \cite{Benvenuti:2011ga}),
\begin{align}
  \lim_{\tau \to +i0} e^{ - \frac{n \pi i}{2 + 4n} \frac{1}{\tau}} \chi_0^{(n)} \sim Z^{S^3}(\xi = i(2n + 1)^{-1}, m) = \int d\sigma \frac{e^{2\pi i \frac{i}{2n + 1}}}{\cosh (\pi \sigma + \frac{m}{2}) \cosh (\pi \sigma - \frac{m}{2})} \ .
\end{align}
Note that the appearance of an imaginary FI-parameter is the hallmark of the mixing of the UV $R$-symmetries with accidental Coulomb branch flavor symmetries \cite{Buican:2015hsa} along the flow from four dimensions to three.\footnote{The paper \cite{Buican:2015hsa} analyzed in detail the case $n=1$.} This is guaranteed to occur whenever the four-dimensional theory contains a Coulomb branch chiral ring operator of non-half-integer $U(1)_r$-charge.

Denoting the twisted translated Higgs branch operator as $Q_{Ai}$, we define the moment map operators of the $\mathfrak{su}(2)$ flavor symmetry
\begin{equation}
  X = \frac{1}{2} Q_{11}Q_{22}, \qquad Y = \frac{1}{2} Q_{12}Q_{21}, \qquad Z = \frac{1}{4}(Q_{11}Q_{21} - Q_{12}Q_{22}) \ .
\end{equation}
Turning off the mass parameters, these operators have vanishing one-point functions, while their two-point and three point functions can be computed with ease following the methods of \cite{Dedushenko:2016jxl},
\begin{equation}
\langle Z(\varphi) Z(0)\rangle = - \frac{(1 + \xi^2 )}{12}\;, \quad  \langle X(\varphi) Y(0)\rangle = - \frac{( 1 + \xi^2)}{6} \;,\quad \langle X(\varphi_1) Y(\varphi_2)Z(0)\rangle = \frac{(1 + \xi^2)}{12} \;.
\end{equation}
The resulting elementary star-products take the form,
\begin{align}
  Z \star Z = & \ (ZZ) \qquad\qquad - \frac{1 + \xi^2}{12} \zeta^2 \\
  Z \star X = & \ (Z X) + \frac{\zeta}{2} X\\
  Z \star Y = & \ (Z Y) - \frac{\zeta}{2} Y \\
  X \star Y = & \ (Z Z) -\zeta Z  \qquad+ \frac{1 + \xi^2}{6} \zeta^2 \ .
\end{align}
As expected, once we substitute in $\xi = \frac{i}{2n + 1}$, this is precisely the same star product as obtained from the VOA. Note that, for example, for $n=1$ the value $\xi=0$ would have reproduced the star-product $\mathbb{DQ}[\text{dim. red. of }(A_1,D_3)]$, see \cite{Beem:2016cbd}. We conclude that the nonzero value of the Fayet-Iliopoulos parameter has modified the period of the quantization.

\subsection{\texorpdfstring{$(A_1, D_{4})$}{(A1,D4)} Argyres-Douglas theories}
The analysis of an arbitrary member of the series of $(A_1, D_{2n+2})$ Argyres-Douglas theories is beyond the scope of this paper, as their associated vertex operator algebras (the so-called $\mathcal{W}_{n + 1}$ algebras \cite{Creutzig:2017qyf}) do not easily lend themselves to the computation of torus correlation functions. However, the $n=1$ member of this series, the $(A_1, D_{4})$ theory, is amenable to a detailed analysis as its associated VOA is simply $\widehat{\mathfrak{su}}(3)_{- 3/2}$ \cite{Beem:2013sza,Cordova:2015nma,Buican:2015ina}. Using the results of appendix \ref{AKM} and the vacuum character
\begin{equation}
\chi_0 = \frac{\eta(2\tau)^8}{\eta(\tau)^8} \ ,
\end{equation}
we immediately find the elementary star product
\begin{align}
J^a \star J^b = (J^a J^b) + \zeta i f^{abc} J^c - \frac{3}{4} \zeta^2 \delta^{ab}\ . \label{su3star}
\end{align}

We can verify this result directly against the computations of the relevant star-products in \cite{Joung:2014qya}. Alternatively, we can use the dimensional reduction of this theory, which is $\mathcal{N} = 4$ SQED with three flavors. Note that in this case there is no mixing between $R$-symmetries and accidental symmetries \cite{Buican:2015hsa}, thus we expect to find the correct period. In fact, as was shown in \cite{Beem:2016cbd}, the minimal nilpotent orbit of $SU(3)$ admits only a unique even, $\mathbb C^*$-equivariant deformation quantization, so there was no room to modify the period to start with. The star product in the three-dimensional setting was computed in \cite{Dedushenko:2016jxl},
\begin{align}
  J_m{^n} \star J_p{^q} = J_{mp}{^{nq}} + \zeta (\delta_p^n J_m{^q} - \delta_m^q J _p{^n}) - \frac{3}{4}\zeta^2 (\delta_m^q \delta_p^n - \frac{1}{3} \delta_m^n \delta_p^q) \ .
\end{align}
Of course it agrees with the results of \cite{Joung:2014qya}, and it also matches \eqref{su3star} by identifying $J^a \equiv J_m{^n} (T^a)^m{_n}$ with properly chosen $\mathfrak{su}(3)$ generators $T^a$ in the fundamental representation, such that $[T^a, T^b] = i f^{abc}T^c$ and $\operatorname{tr}T^a T^b = \delta^{ab}$.

%% file: appendices/JacobiTheta.tex

\section{Special Functions}\label{specialfuctions}

The standard Jacobi theta functions are defined as
\begin{align}
	\vartheta_1(z|\tau) \colonequals & \ -i \sum_{r \in \mathbb{Z} + \frac{1}{2}} (-1)^{r-\frac{1}{2}} e^{2\pi i r z} q^{\frac{r^2}{2}} ,
	& \vartheta_2(z|\tau) \colonequals & \sum_{r \in \mathbb{Z} + \frac{1}{2}} e^{2\pi i r z} q^{\frac{r^2}{2}} \ ,\\
	\vartheta_3(z|\tau) \colonequals & \ \sum_{n \in \mathbb{Z}} e^{2\pi i n z} q^{\frac{n^2}{2}},
	& \vartheta_4(z|\tau) \colonequals & \sum_{n \in \mathbb{Z}} (-1)^n e^{2\pi i n z} q^{\frac{n^2}{2}} \ .
\end{align}
Here we have used the nome $q\colonequals e^{2 \pi i \tau}$. To access their $\tau \to +i0$ behavior, it is useful to note their S-transformation properties,
\begin{equation}
	\vartheta_1\Big( \frac{z}{\tau} | - \frac{1}{\tau}\Big) = - i \alpha(z, \tau)\vartheta_1(z|\tau)\ , \qquad 
	\vartheta_{2,3,4}\Big( \frac{z}{\tau} | - \frac{1}{\tau}\Big) = \alpha(z, \tau)\vartheta_{2,3,4}(z|\tau)\ ,
\end{equation}
where $\alpha(z, \tau) = \sqrt{- i \tau} e^{+ \frac{\pi i z^2}{\tau}}$. As a result we have the high-temperature behavior with $\tau = i \beta$
\begin{align}
  \vartheta_1(z|\tau) \sim \frac{e^{- \frac{\pi}{4\beta}}}{\sqrt{\beta} e^{\frac{\pi z^2}{\beta}}}(2\sinh \frac{\pi z}{\beta} + o(e^{-\frac{2\pi}{\beta}})) \ , \quad
  \vartheta_4(z|\tau) \sim \frac{e^{- \frac{\pi}{4\beta}}(2\cosh \frac{\pi z}{\beta} + o(e^{- \frac{2\pi}{\beta}}))}{\sqrt{\beta} e^{\frac{\pi z^2}{\beta}}} \ .
\end{align}

The Dedekind $\eta$-function is defined as $\eta(\tau) \colonequals q^{\frac{1}{24}}\prod_{n = 1}^{+\infty} (1 - q^n)$. It is related to $\vartheta_1$ through
\begin{equation}
	\vartheta'_1(0|\tau) = 2\pi \eta(\tau)^3\ ,
\end{equation}
and transforms elegantly under the S-transformation
\begin{equation}
	\eta \Big(  - \frac{1}{\tau}  \Big) = \sqrt{-i \tau}\ \eta(\tau) \ .
\end{equation}
We can easily deduce its high-temperature behavior
\begin{equation}
	\eta(\tau) \xrightarrow{\tau \to +i0} \frac{e^{ - \frac{i\pi}{12\tau}}}{\sqrt{-i \tau}} \ .
\end{equation}

Another useful series of functions $\mathcal{P}_m(z|\tau)$ is related to $\vartheta_1(z|\tau)$ as
\begin{equation}
	\mathcal{P}_{m+1}(z|\tau) = - \frac{1}{m!} \partial_z^{m + 1} \ln \vartheta_1(z|\tau) - \pi i \delta_{m,0} \ , \qquad m \ge 0 \ .
\end{equation}
They have high-temperature asymptotics (assuming $\re z > 0$)
\begin{equation}\label{asymptoticOfP}
	\mathcal{P}_1(z|\tau) \xrightarrow{\beta \to +0} \frac{\pi(-1 + 2 \Re z)}{\beta} \ ,
	\quad	\mathcal{P}_2(z|\tau) \xrightarrow{\beta \to +0} \frac{2\pi}{\beta} \ ,
	\quad 	\mathcal{P}_{m > 2}(z|\tau)  \xrightarrow{\beta \to +0} \frac{\left(-\frac{2\pi}{\beta}\right)^m}{(m - 1)!} e^{- \frac{2\pi\Re z}{\beta}}\;. 
\end{equation}
Note the exponential suppression as $\beta \to +0$ for $\mathcal{P}_{m > 2}$.

The Weierstrass $\wp$-function is defined as
\begin{equation}
	\wp(z, \tau) \colonequals - \partial_z^2 \ln \vartheta_1(z|\tau) + 4\pi i \frac{d}{d\tau}\ln \eta(\tau) \ .
\end{equation}
It is an elliptic function with double poles at $z = m + n \tau$ for $m,n\in \mathbb Z$, which is manifest from its alternative definition
\begin{equation}
	\wp(z, \tau) \equiv \frac{1}{z^2} + \sum_{(m, n) \in \mathbb{Z}^2 \backslash (0,0)} \left[\frac{1}{(z + m + n \tau)^2} - \frac{1}{(m + n \tau)^2}\right] \ .
\end{equation}
In particular, its expansion in $z$ around the origin reads
\begin{equation}
	\wp(z, \tau) = \frac{1}{z^2} + 0 + O(z^2) \ ,
\end{equation}
without constant term. The high-temperature behavior of $\wp$ is
\begin{equation}
	\wp(z, i \beta) \xrightarrow{\beta \to +0} \frac{\pi^2}{3 \beta^2}\ , \qquad \re z \ne 0\ .
\end{equation}

To conveniently write down torus correlation functions of current algebras, we also define
\begin{equation}\label{Szegkernels}
	S_i(z |\tau) \equiv \frac{\vartheta'_1(0|\tau)}{\vartheta_i(0|\tau)} \frac{\vartheta_i(z|\tau)}{\vartheta_1(z|\tau)} \ ,
	\qquad
	e_i(\tau) \equiv -4\pi i \frac{d}{d\tau} \ln \frac{\vartheta_i(0|\tau)}{\eta(\tau)} \ .
\end{equation}
The functions $S_i$ are referred to as genus-one Szeg\"{o} kernels \cite{mathur}. They are related to the Weierstrass $\wp$-function by
\begin{equation}
	S_i(z |\tau)^2 = \wp(z, \tau) - e_i(\tau) \ .
\end{equation}
Notice that
\begin{equation}
	\sum_{i = 2,3,4}e_i(\tau) = -4\pi i \frac{d}{d\tau}\ln \frac{\vartheta_2(0|\tau) \vartheta_3(0|\tau) \vartheta_4(0|\tau)}{\eta(\tau)^3} = 0 
\end{equation}
thanks to the standard identity $\vartheta_2(0|\tau)\vartheta_3(0|\tau)\vartheta_4(0|\tau) = 2\eta(\tau)^3$. We thus have
\begin{equation}
\sum_{i = 2,3,4}S_i(z|\tau)^2 = 3 \wp(z,\tau) \ .
\end{equation}

%% file: appendices/AKM.tex

\section{Current algebras and star product}\label{AKM}
In this appendix we present explicit expressions for torus correlation functions of $\hat{\mathfrak g}_k$ currents, derive their high-temperature behavior, and compute the resulting $\star$-products. An affine current algebra $\mathfrak{g}_k$ is defined by the basic operator product expansion
\begin{align}
	J^a(z) J^b(0) = \frac{k\, \kappa^{ab}}{z^2} + \frac{i f^{ab}_{\phantom{ab}c}J^c(0)}{z} + \ldots  \ , \label{currentOPE}
\end{align}
where $f^{ab}_{\phantom{ab}c}$ are the structure constants of the underlying Lie algebra $\mathfrak{g}$, \ie{}, $[T^a, T^b] = i f^{ab}_{\phantom{ab}c}T^c$, and $\kappa^{ab}$ is the Killing form.

Two- and three-point current correlation functions on a torus $\mathbb{T}^2$ with complex modulus $\tau$, for algebras that do not possess a cubic casimir, read\cite{mathur}\footnote{Note that by $\mathfrak{g}$-symmetry $\langle J^a(z)\rangle = 0$.}
\begin{align}
  \langle J^a(z) J^b(0)\rangle
  = & \ k \kappa^{ab} \left[
    \wp(z, \tau)
    + \frac{4\pi i (k+h^\vee)}{k \dim \mathfrak{g}} \frac{d}{d\tau} \ln \chi_0
  \right]\\
  \langle J^a(z_1) J^b(z_2) J^c(0)\rangle =  & \ - i k f^{abc} \sum_{i = 2,3,4}W_i(\tau) S_i(z_1 - z_2 | \tau)S_i(z_2| \tau) S_i(-z_1| \tau) \ .
\end{align}
Here $\chi_0$ denotes the vacuum character of $\widehat{\mathfrak{g}}_k$ and we used the genus-one Szeg\"{o} kernels, see \eqref{Szegkernels}. The $\wp(z,\tau)$ function in the two-point function provides the $z^{-2}$ pole at the origin as mandated by the OPE (\ref{currentOPE}), while the second term accounts for the Sugawara relation
\begin{align}
	\sum_a \lim_{z\to 0} \left[\langle \kappa_{ab} J^a(z)J^a(0)\rangle - \frac{k}{z^2}\right]
	= 2 (k + h^\vee) \langle T(0) \rangle = 4\pi i (k + h^\vee) \frac{d}{d\tau} \ln \chi_0\ .
\end{align}
Here we also used \eqref{stressTensoronept} expressing the one-point function of the stress tensor in terms of the derivative of the vacuum character. The three functions $W_i(\tau)$ are defined by solving the equations 
\begin{align}
	W_2 + W_3 + W_4 = 1, \qquad W_2 e_2 + W_3 e_3 + W_4 e_4 = \frac{4\pi i (k + h^\vee)}{k \dim \mathfrak{g} } \frac{d}{d\tau } \ln\chi_0 \ ,
\end{align}
leaving one immaterial degree of freedom undetermined. To avoid clutter, we have also omitted the subscripts $\mathbb{T}^2$ in $\langle \mathcal{O}\rangle_{\mathbb{T}^2}$. 

To extract a star product, we consider the high-temperature limit of the rescaled correlation functions of the currents. To this end, we assume the Cardy behavior captured by the effective central charge (see also \eqref{Cardy} and \eqref{ceff})
\begin{align}
	\chi_0(q = e^{2\pi i \tau}) \xrightarrow{\tau \to +i0} e^{\frac{2\pi i c_\text{eff}}{24\tau}} (1 + \ldots) \ , \qquad c_\text{eff} \in \mathbb{R} \ ,
\end{align}
which holds for all examples we consider. We also commute the high-temperature asymptotic with $\partial_\tau$ at will. In the end, we find
\begin{align}
	\lim_{\beta \to 0} \beta^2 \langle J^a(z) J^b(0) \rangle = & \ \left(\frac{k}{3} - \frac{(k + h^\vee)c_\text{eff} }{3\dim \mathfrak{g}}\right)\pi^2\kappa^{ab} \ ,\label{JJ2ptf}\\
	\lim_{\beta \to 0} \beta^3 \langle J^a(z) J^b(w) J^c(0) \rangle = & \ \left(\frac{k}{3} - \frac{(k + h^\vee) c_\text{eff}}{3\dim \mathfrak{g}}\right) i \pi^3 f^{abc}\ .\label{JJJ3ptf}
\end{align}
Note that using \eqref{2ptbij2} and \eqref{3ptcijk}, the fact that the two-point and three-point function are controlled by the same combination of quantities could also be seen from the results in appendix C of \cite{Dolan:2007eh}.

Finally, defining $\widetilde{J} \equiv \frac{\zeta}{\pi} J$, the above correlation function leads to the star product
\begin{align}
	\widetilde{J}^a \star \widetilde{J}^b = (\widetilde{J}^a \widetilde{J}^b) + \zeta i f^{abc} \widetilde J^c + \zeta^2 \left(\frac{k}{3} - \frac{ (k + h^\vee)c_\text{eff}}{3 \dim \mathfrak{g}}\right) \kappa^{ab} \ .
\end{align}
We expect this result to remain valid even if the algebra has a cubic casimir. Note that for a current algebra, the central charge is given by the Sugawara central charge,
\begin{align}
	c_\text{2d} = \frac{k \dim \mathfrak{g}}{k + h^\vee} \ ,
\end{align}
while the effective central charge is related to the minimal conformal weight $h_\text{min}$ of the modules appearing in the S-transformation of $\chi_0$ by
\begin{align}
	c_\text{eff} = c_\text{2d} - 24 h_\text{min} \ .
\end{align}
The star product can thus be rewritten as
\begin{align}
	\widetilde{J}^a \star \widetilde{J}^b = (\widetilde{J}^a \widetilde{J}^b) + \zeta i f^{ab}_{\phantom{ab}c} \widetilde J^c + \zeta^2 \frac{8(k + h^\vee)h_\text{min}}{\dim \mathfrak{g}} \kappa^{ab} \ .
\end{align}

%% file: appendices/virasoro.tex

\section{Stress tensor and Virasoro VOA}\label{app:stresstensor}
In this appendix we analyze torus correlation functions of stress tensors making use of Ward identities \cite{Eguchi:1986sb,felder1989}. In particular, we provide an alternative proof that in the high-temperature limit
\begin{align}
	\lim_{\beta\to+0}\beta^{2n}\Big\langle \prod_{i = 1}^n \hat{T}(z_i) \Big \rangle_{\mathbb{T}^2} = 0\ ,
\end{align}
where $\hat{T}(z) \colonequals T(z) - \langle T(z)\rangle_{\mathbb{T}^2}$. We often denote $t = \langle T(z)\rangle_{\mathbb{T}^2}$.

First of all, we note that $\hat{T}$-correlation functions can be reorganized as
\begin{align}
  \Big\langle \prod_{i=1}^n \hat{T}(z_i) \Big\rangle
  = \sum_{p = 1}^n \left[(-t)^{n - p} \sum_{ \{i_1, \ldots, i_p\} } \Big(\big\langle T(z_{i_1}) \ldots T(z_{i_p})\big\rangle - t^p \Big)\right] \ .
  \label{decomposeTcorrelationfn}
\end{align}
Hence, the high-temperature limit of $\langle \prod_{i=1}^n \hat{T}(z_i) \rangle$ can be deduced from the high-temperature behavior of $t^{n - p}$ and the combinations $\langle \prod_{i=1}^p {T}(z_i) \rangle - t^p$ with $p \le n$. The latter piece of information can be extracted by carefully analyzing the relation \cite{felder1989} (slightly rearranged as compared to that reference)
\begin{align}
  & \ \Big\langle T(z) \prod_{j=1}^p T(z_j)\Big\rangle - t^{p + 1}\\
  = & \ 2\pi i \frac{1}{\chi_0} \frac{d}{d\tau} \bigg(\chi_0 \Big[\big\langle \prod_{j=1}^p T(z_j) \big\rangle -t^p\Big] \bigg) + 2\pi i \frac{d}{d\tau} t^p \nonumber\\
    & \ + \sum_{j = 1}^p \left[
      2\big( \wp(z - z_j) + 2 \eta_1 \big)
      + \big(  \zeta(z - z_j) + 2\eta_1 z_j  \big) \frac{\partial}{\partial z_j} 
    \right]\bigg[\Big\langle \prod_{j=1}^p T(z_j)\Big\rangle - t^p\bigg]\label{recursion1} \\
    & \ + \frac{c}{12} \sum_{j = 1}^p \wp''(z - z_j) \bigg[\Big\langle \prod_{\substack{i=1 \\ i \ne j}}^p T(z_i)\Big\rangle - t^{p - 1}\bigg]
    + 2t^p\sum_{j = 1}^p \big( \wp(z - z_j) + 2\eta_1 \big)
    + \frac{c t^{p - 1}}{12} \sum_{j = 1}^p \wp''(z - z_j)\nonumber
\end{align}
Here $\eta_1$ and $\zeta(z)$ are defined as
\begin{equation}
  \eta_1 \colonequals - \frac{2 \pi i}{3} \frac{\partial_\tau \vartheta_1'(0|\tau)}{\vartheta_1'(0|\tau)}\ , \qquad \zeta(z) \colonequals \frac{\vartheta_1'(z|\tau)}{\vartheta_1(z|\tau)} + 2 z \eta_1 \ .
\end{equation}

It is straightforward to work out the high-temperature behavior of the various building blocks of (\ref{recursion1}) when $\re z \ne 0$, $\re (z - z_1) > \re z_1 > 0$
\begin{align}
  & \wp(z) + 2 \eta_1 \xrightarrow{\beta \to 0}\frac{2\pi}{\beta} \ ,
  \qquad
  \ \zeta(z - z_1) - 2z_1 \eta_1 \xrightarrow{\beta\to0} - \frac{2\pi^2 \Re z}{3 \beta^2}, \\
  & \wp''(z) \xrightarrow{\beta \to 0} \frac{16\pi^4 e^{-\frac{2\pi |\Re z|}{\beta}}}{\beta^4} \ ,
\end{align}
while the vacuum character $\chi_0$ exhibits Cardy behavior as before.

To proceed, we allow ourselves to pass $\partial_z$ and $\partial_\tau$ acting on the correlation functions past taking the leading term in the high temperature limit with impunity.\footnote{For correlation functions in VOAs associated with Lagrangian SCFTs, one can explicitly verify that these actions indeed commute.} First off, $\langle T(z)\rangle - t$ vanishes identically. Next we have, requiring $\re (z - z_1) > 0$,
\begin{align}
  \langle T(z) T(z_1) \rangle - t^2
  = & \ \left(2\pi i \frac{d}{d\tau} \ln \chi_0\right) t + 2\pi i \frac{d}{d\tau} \langle T(z_1)\rangle - t^2 \nonumber \\
    & \ + 2 \big(  \wp(z-z_1) + 2\eta_1  \big) t  + \big(  \zeta(z - z_j) + 2 \eta_1 z_1 \big) \partial_{z_1} t \\
    & \ + \frac{c}{12} \wp''(z - z_1) \nonumber \ .
\end{align}
The first and last term in the first line cancel by the relation between $t$ and $\ln \chi_0$. The remaining terms exhibit high-temperature behavior (here we have dropped irrelevant numerical factors to avoid clutter)
\begin{align}
  & t = \langle T(z_1)\rangle = 2\pi i \frac{d}{d\tau}\Big|_{\tau = i \beta} \ln \chi_0 \xrightarrow{\beta \to +0} \frac{1}{\beta^2} , \qquad \qquad 2\pi i \frac{d}{d\tau} \langle T(z_1)\rangle \xrightarrow{\beta \to +0} \frac{1}{\beta^3}\ ,\nonumber \\
  & (\wp(z - z_1) + 2\eta_1) t \xrightarrow{\beta \to +0} \frac{1}{\beta^3}, \qquad \qquad \qquad\qquad \quad \ (\zeta(z- z_1) - 2z_1)\partial_{z_1}t = 0 \ ,\\
  & \wp''(z - z_1) \xrightarrow{\beta \to +0} \frac{e^{- \frac{2\pi}{\beta} |\Re (z - z_1)|}}{\beta^4} \ . \nonumber
\end{align}
Apparently, terms in $ \langle T(z) T(z_1) \rangle - t^2$ either have at most $\beta^{-3} = \beta^{- (2 \times 2 - 1)}$ singularity or are exponentially suppressed; once rescaled by $\beta^4$, the correlation function vanishes as $\beta \to +0$.

This observation, combined with (\ref{recursion1}), initiates a recursive argument ensuring that higher correlation functions $\beta^{2(p+1)}(\langle T(z) \prod_{j=1}^pT(z_j)\rangle - t^{p + 1})$, for $p \ge 2$, also vanish in the high-temperature limit. Indeed, assuming the high-temperature behavior of $\langle \prod_{j=1}^p T(z_j)\rangle - t^p$ has at most $\beta^{-(2p - 1)}$ singularity, terms on the right hand side of (\ref{recursion1}) either have at most a $\beta^{- (2p + 1)}$ singularity, like $\partial_\tau t^p$, or are exponentially suppressed, like the $\wp''(z - z_j)$ terms. Once rescaled by $\beta^{2p + 2}$, they go away in the high-temperature limit. Finally, this observation is to be applied back to (\ref{decomposeTcorrelationfn}), which clearly shows that $\beta^{2n}\langle \prod_{i=1}^{n} \hat{T}(z_i)\rangle$ vanishes in the same limit. Furthermore, inserting spatial derivatives or forming composites of $T$ by taking the regular part of coincident limits will not make the correlation functions nonzero.

To conclude, the set of correlation functions only involving the stress-energy tensor decouples. Moreover, slightly modified, we expect the above argument to extend to all correlation functions containing a stress energy tensor.

%% file: appendices/characters.tex

\section{Characters}\label{characters}
In this appendix we collect characters of various VOAs used in the main text.

The Virasoro minimal models $\operatorname{Vir}_{p,p'}$ are labeled by two coprime natural numbers $p, p' \ge 2$ with central charge $c = 1- 6 \frac{(p - p')^2}{p p'}$. The spectrum of $\operatorname{Vir}_{p,p'}$ contains a finite number of primaries of conformal dimensions $h_{r,s} = \frac{(p r - p's)^2 - (p - p')^2}{4 p p'}$, $1 \le r < p', 1 \le s < p$, each parenting a degenerate Virasoro representation. The characters of these representations are given by
\begin{align}
	\chi_{r,s}(\operatorname{Vir}_{p,p'}) = \frac{1}{\eta(\tau)} \bigg[
	  q^{\frac{\lambda_{r,s}^2}{4p p'} }\vartheta_3(\lambda_{r,s}\tau|2p p' \tau)
	  - q^{\frac{\lambda_{r, - s}^2}{4p p'} }\vartheta_3(\lambda_{r, - s}\tau|2p p' \tau)
	\bigg]\ ,
\end{align}
where $\lambda_{r,s} \equiv pr - p's$. For each $\operatorname{Vir}_{p,p'}$, $\chi_{r= 1, s = 1}$ is the vacuum character.

The current algebra $\widehat{\mathfrak{su}(2)}_{-1/2}$ is the $\mathbb{Z}_2$-orbifold of the $\beta \gamma$ system, the latter being the associated VOA of a four-dimensional free hypermultiplet. The (flavored) vacuum character of $\widehat{\mathfrak{su}(2)}_{-1/2}$ is given by \cite{Ridout:2008nh}
\begin{align}
	\chi(\widehat{\mathfrak{su}(2)}_{-1/2}) = \frac{1}{2} \left(\frac{\eta (\tau)}{\vartheta_3(\mathfrak{a}|\tau)} + \frac{\eta (\tau)}{\vartheta_4(\mathfrak{a}|\tau)}\right)\ .
\end{align}

The (flavored) vacuum characters of $\widehat{\mathfrak{su}(2)}_{- \frac{4n}{2n+1}}$ series associated with the $(A_1, D_{2n+1})$ Argyres-Douglas theories are given by \cite{Cordova:2015nma}
\begin{align}
	\chi \Big(\widehat{\mathfrak{su}(2)}_{- \frac{4n}{2n+1}} \Big)(\mathfrak{a})
	= \frac{\vartheta_1(\mathfrak{a}|(2n + 1)\tau)}{\vartheta_1(\mathfrak{a}|\tau)} \xrightarrow{\mathfrak{a} \to 0}  \frac{\eta( (2n + 1)\tau)^3}{\eta(\tau)^3}\ .
\end{align}
With the S-transformation properties summarized in appendix \ref{specialfuctions}, it is easy to extract the Cardy behavior
\begin{align}
	\ln \chi \Big(\widehat{\mathfrak{su}(2)}_{- \frac{4n}{2n+1}} \Big)(\mathfrak{a})
	\to \frac{\pi i n }{4n + 2} \frac{1}{\tau} = 4\pi i (c_\text{4d} - a_\text{4d}) \frac{1}{\tau} \ ,
\end{align}
for the central charges of $(A_1, D_{2n+1})$ theories
\begin{align}
	c_\text{4d} = \frac{n}{2}, \qquad a_\text{4d} = \frac{8n^2 + 3n}{16n + 8} \ .
\end{align}

Four-dimensional $\mathcal{N} = 4$ SYM with gauge group $SU(2)$ is associated to the two-dimensional small $\mathcal{N} = 4$ superconformal algebra at central charge $c=-9$. The vacuum character can be computed by the contour integral
\begin{align}
  \chi_0 = \frac{1}{2} \oint \frac{da}{2\pi i a} \frac{\vartheta_1(2\mathfrak{a}|\tau) \vartheta_1(-2\mathfrak{a}|\tau) \eta(\tau)^3}{\vartheta_4(2\mathfrak{a}|\tau) \vartheta_4(-2\mathfrak{a}|\tau) \vartheta_4(0|\tau)} = q^{\frac{3}{8}} + \chi^{\mathfrak{su}(2)}_3(a^2) q^{\frac{11}{8}} - 2 \chi^{\mathfrak{su}(2)}_2(a^2) q^{\frac{15}{8}} + \ldots \ . \nonumber
\end{align}
Its leading Cardy behavior can be accessed by analyzing its integrand, and as expected the leading $\tau^{-1}$-Cardy behavior vanishes.

Finally, the vacuum character of $\widehat{\mathfrak{su}(3)}_{-3/2}$ is given by \cite{Arakawa:2016hkg}
\begin{align}
	\chi_0 = \frac{\eta(2\tau)^8}{\eta(\tau)^8} \ ,
\end{align}
leading to the Cardy behavior $\ln \chi_0 \to \frac{i\pi}{3\tau}$.